\documentclass{jfm}

\usepackage{here}
\usepackage{float}
\usepackage{amsmath}
\usepackage{wrapfig}
\usepackage{makecell}
\usepackage[utf8]{inputenc}
\usepackage{multirow}
\usepackage{verbatim}
\usepackage{mathrsfs}
\usepackage{comment}
\usepackage{caption}
\usepackage{subcaption}
\usepackage[percent]{overpic}
\usepackage{graphicx}
\usepackage{newtxtext}
\usepackage{newtxmath}
\usepackage{natbib}
\usepackage{hyperref}
\hypersetup{
    colorlinks = true,
    urlcolor   = blue,
    citecolor  = black,
}

\newcommand{\RomanNumeralCaps}[1]

\title{Micro-convection mass transfer following bubble coalescence on a solid wall}

\author{Zahra Saadat\aff{1}\footnote{\label{note1}These authors contributed equally to this work and share first authorship.},
Michele Cattani\aff{1}$\dagger$,
Amin Soltani\aff{1}
\and Pourya Forooghi\aff{1}
 \corresp{\email{forooghi@mpe.au.dk}}}

\affiliation{\aff{1}Department of Mechanical \& Production Engineering, Aarhus University, 8200 Aarhus, Denmark
}

\begin{document}
\maketitle

\begin{abstract}
The Volume-of-Fluid implementation in the \href{basilisk.fr}{Basilisk} flow solver is employed to study mass transfer after coalescence-induced jump-off of bubbles on solid substrates at four combinations of bubble radius ($R_m=150$ and 25 $\mu$m, hydrogen-water properties) and Schmidt number ($\text{Sc}=210$ and 1). The results show a relatively strong downward entrainment of low-concentration liquid, induced by a rapid movement of the bubble interface at approximately 0.5 to 1 inertio-capillary time units after the moment of coalescence. At $\text{Sc}=210$, this leads to a highly local increase of the Sherwood number at a core region (roughly an area of radius $R_m/2$) below the south pole of the merged bubble, which persists long after the bubble departure due to the slow diffusion at large $\text{Sc}$. The enhancement factor of the Sherwood number is highly dependent on the state of the mass-transfer boundary layer at the moment of coalescence, increasing with smaller boundary layer thicknesses. Shortly after the jump-off, the velocity of the bubble is considerably damped and rapidly approaches its free-rise terminal velocity. The effect of micro-convection on the wall mass transfer coefficient at this stage is insignificant when isolated, similar to what is reported in the literature for purely buoyancy-driven bubble rise -- but slightly stronger.

\end{abstract}


{\bf MSC Codes }  {\it(Optional)} Please enter your MSC Codes here


\section{Introduction}
\label{sec:headings}

Several industrial applications, e.g., boiling and electrolysis, involve formation and departure of gas bubbles on solid walls, where wall heat or mass transfer is at the core of the process. Specifically, the effect of bubbles on wall mass transfer in the context of gas-evolving electrochemical processes has been a research topic for several decades \citep{janssen1970effect,stephan1979model,janssen1985mechanism,taqieddin2017physicochemical,valle2024analytical}. While both the mass transfer between wall and liquid and that between liquid and bubbles can be of importance, the present work is focused on the former. The classical view is that bubbles can enhance mass transfer at the micro-scale through three mechanisms \citep{vogt2015local}: (1) growth of a bubble pushes the surrounding liquid, causing convection; (2) bubble departure off the solid surface leads to convection in its wake; and (3) a detached bubble disturbs the concentration boundary layer behind it, leading to additional mass diffusion. The first two mechanisms are commonly referred to as `micro-convection'. The present work primarily concerns the second mechanism.

Empirical evidence confirms the enhancement of mass transfer due to bubble generation, with results often pointing towards a power-law relation between the mass transfer coefficient and the rate of gas production \citep{fouad1973mass,bockris1980comprehensive,el1991mass}. The detailed contributions of different mechanisms and the exact value of the power-law exponent have, however, been a matter of debate in the literature. The earliest attempt to formulate the problem was made by \cite{ibl1971stofftransport}, who mainly focused on diffusive mass transfer after bubble departure. Later, \cite{stephan1979model} and \cite{vogt2015local} developed other mass transfer models with special focus on the effect of micro-convection. Meanwhile, \cite{janssen1970effect,janssen1985mechanism} hold the view that the main cause of mass transfer enhancement with gas production rate is the buoyancy-driven mixing of the concentration field. This mechanism is more commonly referred to as 'macro-convection' because it involves mixing in the bulk of the liquid. Experiments by \cite{janssen1979effect} confirmed the theoretically expected $1/3$ power law exponent for this mechanism when bubble coalescence is absent, but they observed significant deviation when coalescence events occur frequently. Recently, \cite{sepahi2024mass} revisited the problem using direct numerical simulations of periodic buoyancy-driven bubble departure from a horizontal wall, and reported the same scaling law. These authors also did not account for bubble coalescence. 

An isolated gas bubble is detached from a horizontal surface due to the action of the buoyancy force when its diameter reaches a certain threshold commonly known as the `Fritz' diameter, following the work of \cite{fritz1935berechnung}. In real systems, however, bubbles may be detached at smaller radii as a result of their coalescence, among other effects. When two bubbles (or similarly two droplets) coalesce and merge on a solid wall, the merged bubble has a smaller interface area compared to the sum of the two 'parents' when reaching its equilibrium spherical shape. This difference in the surface area leads to a release of surface potential energy and subsequently a 'jump-off' effect \citep{liu2014numerical,soto2018coalescence}. The merged bubble jumps off the wall, irrespective of the buoyancy force, as long as its size is larger than a certain threshold at the time of coalescence and the contact angle is smaller than a certain size-dependent value \citep{iwata2022coalescing,demirkir2025jump}. Notably, \cite{bashkatov2022growth} experimentally confirmed that bubbles can depart an electrode surface even under micro-gravity conditions.

The autonomous jump-off was first reported by \cite{boreyko2009self} for droplets on super-hydrophobic surfaces. Subsequently, several authors used analytical and experimental methods to investigate the dynamics of droplet \citep{wang2011size, wang2018morphology} or bubble \citep{soto2018coalescence,bashkatov2024performance,zhang2024coalescence} coalescence on solid substrates. One of the first interface-resolving numerical simulations of droplet coalescence was carried out by \cite{liu2014numerical}, who elucidated, among others, the balance of released surface potential energy with the kinetic energy and viscous dissipation following the coalescence. Later on, several authors studied different aspects of droplet coalescence using numerical simulations  \cite[see, e.g.,][]{chen2018numerical,wang2019dynamic, chen2019numerical}. For bubble coalescence, the number of numerical studies is more limited: \cite{iwata2022coalescing,zhao2022coalescence} focused on developing regime maps for the occurrence of jump-off in large bubbles, for which the contact angle is the determining factor. Meanwhile, \cite{cattani2026study} investigated the dynamics and energetics of bubble coalescence for smaller bubbles, for which the viscous dissipation is determining. In oddition to these numerical efforts, certain authors have developed jump-off regime maps based on experimental data and scale analysis \citep{lv2021self,demirkir2025jump}. All the above-mentioned works on droplet and bubble coalescence focused primarily on the hydrodynamic aspects and not on the scalar transport. 

Mass transfer between a bubble and surrounding liquid has been the subject of several theoretical and experimental studies in the past, focusing on either isolated bubbles or those growing on solid surfaces \citep{epstein1950stability,scriven1995dynamics,glas1964measurements, jones1999bubble, li2014growth,enriquez2014quasi,penas2016history,penas2017history,van2017electrolysis}. More recently, interface-resolving numerical simulations have been used to provide a more detailed picture. Among others, \cite{maes2018new,vachaparambil2020modeling,gennari2022phase} developed interfacial mass transfer models within the Volume of Fluid (VoF) method, enabling study of bubble growth in electrochemical applications. Meanwhile, \cite{farsoiya2023direct} used their mass transfer model implemented in the Basilisk solver \citep{popinet2009accurate,popinet2015quadtree} to investigate dissolution of a bubble in isotropic homogeneous turbulence. Meanwhile, \cite{han2025numerical} developed a model in Basilisk allowing study of the dynamic contact-line effects on the growing bubbles. Furthermore, \cite{huang2025numerical} numerically examined the growth of wall-attached bubbles with contact lines initially `pinned' to the edge of a micro-cavity.

While the above studies mostly concern bubble growth and liquid-gas mass transfer, \cite{sepahi2024mass} paid special attention to the `wall' mass transfer. They investigated the mass transfer during and after departure of spherical bubbles on a horizontal wall and concluded that buoyancy-driven mixing dominates over micro-convection.  Previously, \cite{sepahi2022effect} also reported a non-negligible solutal convection effect prior to bubble departure. Furthermore, \cite{khalighi2023hydrogen} focused on pre-departure mass transfer near spherical bubbles in realistic conditions corresponding to an alkaline electrolyzer cathode and showed that a cross-flow can moderately increase the wall mass transfer rate. Recently, \cite{qin2026three} studied growth and departure of bubbles on an electrode considering a single as well as multiple nucleation sites. These authors used a VoF implementation in Basilisk, which accounts for bubble deformability and thereby enables the study of the merging phase of neighboring bubbles. They observed that when bubbles merge, it moderately advances the departure and potentially improves the mass transfer. The main bubble attachment mechanism in this work is deemed to be buoyancy rather than coalescence based on the reported Fritz-radius alignment. Meanwhile, \cite{vachaparambil2021numerical} employed a geometric VoF method to study coalescence-driven departure of bubbles from solid electrodes with mass transfer. These authors employed a 2D setup and mainly focused on the coalescence dynamics rather than wall mass transfer.

Overall, limited attention has been paid in the literature to wall mass transfer induced by bubble coalescence. This leaves unanswered questions, specifically when it comes to a complete understanding of micro-convection. While there is some consensus in the recent literature that the micro-convection effects are minor when bubbles detach due to buoyancy, bubble coalescence can cause significant jump-off velocities even for bubbles much smaller than the Fritz diameter \citep{soto2018coalescence,cattani2026study}, which can be a potential cause of strong micro-convection. The present research is mainly aimed at studying this effect. Note that, as will be discussed in the body of the paper, the Reynolds number based on bubble jump-off velocity is comparable to the inverse Ohnesorge number, which is of the order of 50 for a bubble as small as 25 $\mu$m. Motivated by that, in the present work, we employ 3D interface-resolving simulations using the VoF implementation in Basilisk \citep{popinet2009accurate,popinet2015quadtree,farsoiya2021bubble} to study mass transfer following coalescence of two similar bubbles at a Schmidt number corresponding to that of dissolved hydrogen in water. Due to the high computational cost and the intrinsic short time scale of coalescence, we adopt a simplified generic setup, with simulations starting at the moment of coalescence rather than considering the entire lifetime of a growing bubble and its mass-transfer history. Moreover, to isolate the micro-convection effect, factors such as solutal convection and contact-line dynamics were not considered. The objective is, therefore, to shed light on the potential significance of micro-convection mass transfer induced by bubble jump-off.

The paper is organized as follows: in section \ref{sec:methodology}, details of the problem setup and numerical solution are described. Section \ref{sec:results} contains the main results, starting with a description of bubble and surrounding liquid velocities, followed by presentation of the results for concentration and wall mass transfer, and lastly, a discussion of how the results should be interpreted in real-world scenarios. Section \ref{sec:conclusion} summarizes the main findings.


\section{Methodology}
\label{sec:methodology}

\subsection{Problem overview}
\label{sec:problem}
\begin{figure}[H]
\centering

\begin{subfigure}[t]{0.5\textwidth}
    \raggedright $(a)$
    \vspace{2mm}

    \centering
    \begin{overpic}[width=\textwidth]{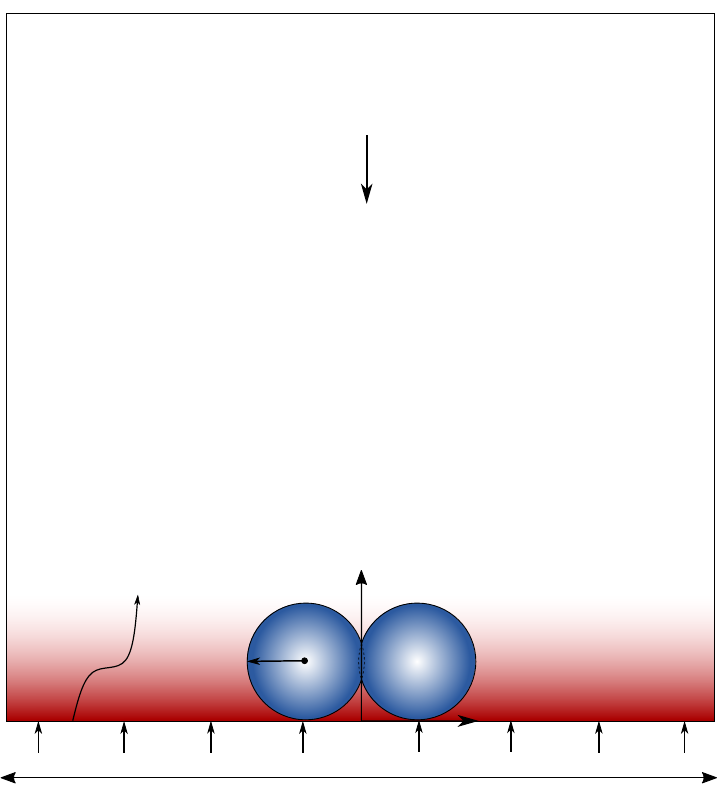}
    \fontsize{10}{11}\selectfont

        \put(50,102){\makebox(0,0){$\partial C/\partial y=0$}}
        \put(50,92){\makebox(0,0){$C=C_\mathrm{b}=0$}}

        \put(50,80){\makebox(0,0){$g$}}

        \put(-5,48){\rotatebox{90}{$\partial C/\partial x=0$}}

        \put(92,48){\rotatebox{90}{$\partial C/\partial x=0$}}

        \put(20,27){\makebox(0,0){$\mathbf{u}=0,\; \frac{\partial C_l}{\partial y}=\frac{J}{\mathscr{D}_1}$}}

        \put(38,19){\makebox(0,0){$R_0$}}

        \put(49,26){$y$}
        \put(60,10){$x$}

        \put(44,3){$J$}

        \put(46,-2){\makebox(0,0){$24R_m$}}

    \end{overpic}

\end{subfigure}
\hspace{0.05\textwidth}
\begin{subfigure}[t]{0.36\textwidth}
    \raggedright $(b)$
    \vspace{2mm}

    \centering
    \raisebox{1.25cm}{%
    \begin{overpic}[width=\textwidth]{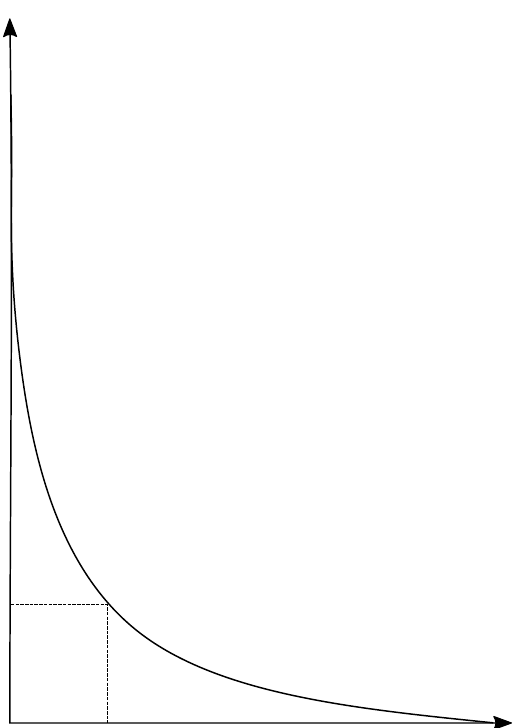}
    \fontsize{10}{11}\selectfont
    
        \put(5,95){\makebox(0,0){$y$}}
        
        \put(68,-5){\makebox(0,0){$C^{1\text{D}}$}}
        
        \put(26,-5){\makebox(0,0){$C_i=0.1C_{\mathrm{w},i}$}}

        \put(-2,15){\makebox(0,0){$\delta_i$}}

    \end{overpic}
    }
\end{subfigure}
  \caption{Schematic of the computational setup. (a) Computational domain and boundary conditions. (b) Initial dissolved-gas concentration field based on a one-dimensional concentration profile near the substrate.}
\label{fig:schematic}
\end{figure}
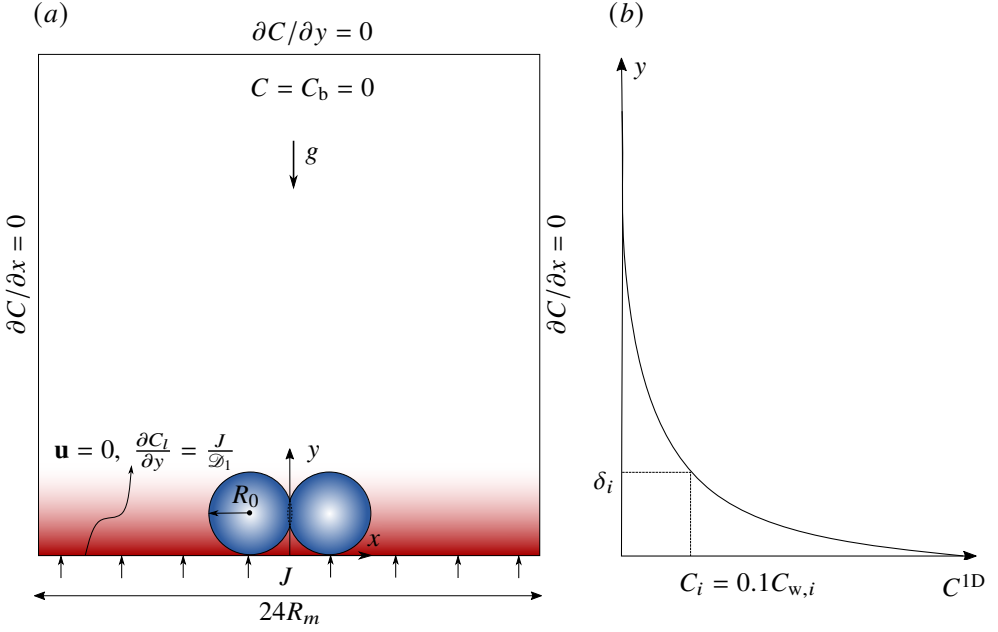

We solve the two-phase flow and mass transfer in a cubic (square in 2D) domain, where two bubbles of radius $R_0$, placed tangentially on the wall, merge at time $t=0$. The setup is shown schematically in Figure \ref{fig:schematic}a. To initiate the merger of the bubbles, we impose an initial overlap of $0.02R_0$, which, as shown in \citep{cattani2026study}, has a negligible impact on the dynamics of the bubble after merger. The radius of the merged bubble is therefore $R_m=2^{1/3}R_0$ ($R_m=2^{1/2}R_0$ in 2D). To compare bubble departure driven by coalescence and buoyancy, additional simulations are conducted in which the rise of a single bubble of radius $R_m$ is studied (not shown in the schematic). Throughout this work, species concentration in the liquid and gas phases is denoted by $C_l$ and $C_g$, respectively. Additionally, the density, dynamic viscosity, kinematic viscosity, diffusivity, surface tension coefficient, and dimensionless solubility coefficient are denoted by $\rho$, $\mu$, $\nu$, $\mathscr{D}$, $\sigma$, and $\alpha$, respectively. At the wall, a constant mass flux of $J=\mathscr{D}_l(\partial C_l/\partial y)_{y=0}$ is prescribed, while the concentration far from the wall, $C_\mathrm{b}$, is virtually zero for the duration of the simulations. We prescribe no-slip and fully wet conditions for the velocity and volume-fraction fields at the wall, respectively, whereas symmetry boundary conditions are imposed on the lateral and top boundaries. Under the fully wet assumption, no three-phase contact line is considered. Although real bubbles always exhibit a finite contact angle, this effect is neglected here as a simplifying assumption to avoid over-complicating this first study.

The simulations are performed over a range of bubble sizes and transport properties. We consider merged bubbles of radii $R_m=25$ $\mu\mathrm{m}$ and $R_m=150$ $\mu\mathrm{m}$. The smaller radius is close to the threshold where the viscous dissipation prevents coalescing bubbles from jumping \citep{cattani2026study}. For the buoyancy-driven problem, only the larger bubble size is studied. Schmidt numbers, $\mathrm{Sc}=\nu_l/\mathscr{D}_l$, of 1, 5, 30, and 210 are considered -- the largest value corresponding roughly to the properties of hydrogen and water. Due to the high computational cost, only a limited number of cases are run in 3D, and intermediate Schmidt values are considered only in the 2D simulations. The simulated cases are summarized in Table~\ref{tab:cases}.

\begin{table}
\centering
\renewcommand{\arraystretch}{1.25}
\begin{tabular}{c@{\hspace{3em}}c@{\hspace{3em}}c@{\hspace{3em}}c@{\hspace{3em}}c}

Type & $R_m\,(\mu\mathrm{m})$ & Sc & $\delta_i/R_m$ & Mechanism \\
\hline

\multirow{4}{*}{3D}
 & \multirow{2}{*}{150} & 210   & \multirow{4}{*}{1/3} & \multirow{4}{*}{Coalescence} \\
 & & 1 & & \\
 & \multirow{2}{*}{25} & 210   & & \\
 & & 1 & & \\

\hline
{Axisymmetric} & {150} &  210 & {1/3}
& {Buoyancy} \\
\hline

\multirow{7}{*}{2D}

& \multirow{7}{*}{150}
& \multirow{4}{*}{210}
& 1/5
& \multirow{7}{*}{Coalescence} \\
& & & 1/3 & \\
& & & 1 & \\
& & & 2 & \\

& & \multirow{1}{*}{30}
& 1/3 & \\

& & \multirow{1}{*}{5}
& 1/3 & \\

& & \multirow{1}{*}{1}
& 1/3 & \\

\hline
\end{tabular}
\caption{
Summary of the simulation cases.}
\label{tab:cases}
\end{table}

In addition to the Schmidt number and the solubility constant $\alpha$, the problem is defined by five dimensionless numbers: the Ohnesorge number, $\mathrm{Oh}=\frac{\mu_l}{\sqrt{\rho_l \sigma R_m}}$; the Bond number, $\mathrm{Bo}=\frac{(\rho_l-\rho_g)gR_m^2}{\sigma}$, and the property ratios, $\rho_l/\rho_g$, $\mu_l/\mu_g$, and $\mathscr{D}_l/\mathscr{D}_g$. Using the properties of hydrogen and water, a 150 $\mu$m bubble corresponds roughly to $(\mathrm{Oh},\mathrm{Bo})=(0.0086,0.003)$ and a 25 $\mu$m bubble to $(\mathrm{Oh},\mathrm{Bo})=(0.021,8\times10^{-5})$. Additionally, the property ratios are fixed at $\mathscr{D}_l/\mathscr{D}_g=10^{-4}$, $\rho_l/\rho_g=1000$, and $\mu_l/\mu_g=103$, which approximately correspond to the hydrogen--water system. Only for the density ratio, we use a smaller value to avoid extreme numerical stiffness; the effect of a density ratio in this range is commonly considered negligible.

The initial concentration field in the liquid is obtained from the solution of the one-dimensional pure diffusion into a semi-infinite domain with Neumann boundary conditions. This is a textbook problem whose solution, for the time $\tilde{t}$ after the onset of diffusion into a semi-infinite domain with an initial concentration of zero, is given by \citep{crank1979mathematics}:
\begin{equation}\label{eq:1D}
C^{1\text{D}} = C_\mathrm{w}^{1\text{D}}\left[\exp(-\hat{y}^2)
- \sqrt{\pi}\,\hat{y}\,\mathrm{erfc}(\hat{y})\right] \quad,\quad C_\mathrm{w}^{1\text{D}}=\frac{2}{\sqrt{\pi}}\sqrt{\frac{\tilde{t}}{\mathscr{D}_l}}J
\end{equation}
where $\hat{y}=y/\sqrt{4\mathscr{D}_l\tilde{t}}$ is the normalized distance from the boundary and erfc$(\square)$ is the complementary error function. Furthermore, we define a diffusion layer thickness, $\delta$, which is the wall distance at which $C^{1\text{D}}$ is 10\% of the wall concentration $C^{1\text{D}}_\mathrm{w}$ (see Figure~\ref{fig:schematic}b). It can be derived from equation \ref{eq:1D} that $\delta=1.92\sqrt{\mathscr{D}_l\tilde{t}}$. In the present study, we prescribe the thickness of the diffusion layer at the moment of coalescence, $\delta_i$, as an input parameter from which the time $\tilde{t_i}$ can be derived. In other words, the gas bubbles are superimposed on the concentration field at time $\tilde{t_i}$ after the beginning of species diffusion into the domain (or $t=\tilde{t}-\tilde{t_i}$). In a real-world scenario, diffusion before coalescence is not one-dimensional, and bubble growth and mass diffusion take place simultaneously. Therefore, the concentration field at the moment of coalescence can be determined by transient simulation of several cycles of bubble growth and departure, similar to the study of \cite{sepahi2024mass}, until a dynamic steady state is reached. However, such a simulation in the present setup would lead to a prohibitively high computational cost. We therefore opted for a more generic initial condition to provide a first insight into the details of post-coalescence convection. Notably, the results of \cite{sepahi2024mass} show that, at least at large current densities, the concentration field around a departing bubble closely resembles that of one-dimensional diffusion once the dynamic steady state is reached. Furthermore, the present setup allows us to vary the initial thickness of the diffusion layer independent of the bubble radius. This is consistent with the fact that bubble coalescence is a random event and does not necessarily occur at a fixed state of the boundary layer.
Indeed, the possibility to systematically vary $\delta_i/R_m$ provides further insight into the interaction of jumping bubbles and the diffusion layer under different scenarios. Once we prescribe $\delta_i$, the value of $\tilde{t_i}$ can be determined from $\delta_i=1.92\sqrt{\mathscr{D}_l\tilde{t_i}}$ and used to prescribe the initial concentration profile $C^\text{1D}/C^\text{1D}_\mathrm{w}$ based on equation \ref{eq:1D}. Furthermore, we assign an initial concentration of unity inside the bubble and an initial liquid concentration $C_{\mathrm{w},i}$ at the wall. The latter value determines $J$ according to equation \ref{eq:1D}. Related to the wall concentration, one can define a quantity $\xi_\mathrm{w}=C_{\mathrm{w}, i}/\alpha-1$, which can be considered the super-saturation of the liquid phase at the wall if the gas inside the bubble is assumed saturated. In the following, we simply refer to $\xi_\mathrm{w}$ as the initial wall supersaturation, and our results include a study of the effect of $\xi_\mathrm{w}$ to measure the sensitivity to the prescribed level of supersaturation.

In this study, we frequently use the inertio-capillary time and velocity scales for normalization, which are defined as
\begin{equation}
t_{ic} = \sqrt{\frac{\rho_l R_m^3}{\sigma}},
\qquad
u_{ic} = \sqrt{\frac{\sigma}{\rho_l R_m}}.
\end{equation}
Additionally, certain quantities are normalized based on time and velocity scales driven by gravitational acceleration and bubble radius, namely buoyancy scales:

\begin{equation}
t_{b}=\sqrt{\frac{R_m}{g}},
\qquad
u_{b}=\sqrt{gR_m}.
\end{equation}

Once the simulations run, the local rate of mass transfer from the wall into the liquid can be quantified through the local Sherwood number,
\begin{equation}
\mathrm{Sh}(x,z,t)=\frac{JR_m}
{\mathscr{D}_l \left(C_\mathrm{w}(x,z,t)-C_\mathrm{b}\right)}.
\end{equation}
where $C_\mathrm{w}(x,z,t)$ denotes the time-dependent local wall concentration obtained from the simulations. To provide a direct quantification of the effect of micro-convection, the Sherwood number is normalized using the corresponding one-dimensional purely diffusive solution in a semi-infinite domain at the corresponding time,
\begin{equation}
\mathrm{Sh}^{1\text{D}}(t)=\frac{JR_m}{\mathscr{D}_l \left(C_\mathrm{w}^{1\text{D}}(t)-C_\mathrm{b}\right)}.
\end{equation}

The normalized Sherwood number is denoted by an asterisk,
\begin{equation}
\mathrm{Sh}^*(x,z,t)=\frac{\mathrm{Sh}(x,z,t)}{\mathrm{Sh}^{1\text{D}}(t)},
\end{equation}
and similarly, $C^*(x,y,t)$ denotes local concentration normalized by $C_\mathrm{w}^{1\text{D}}(t)$.

\subsection{Numerical solution}

We solve the incompressible two-phase flow using the open-source solver Basilisk (\href{http://basilisk.fr}{basilisk.fr}). This includes the solution of the mass and momentum balances,
\begin{gather}
\nabla \cdot \mathbf{u} = 0,
\\
\rho \left(\partial_t\mathbf{u}+
\mathbf{u}\cdot\nabla\mathbf{u}\right)=-\nabla p+\nabla\cdot\left[\mu\left( \nabla\mathbf{u}+\nabla\mathbf{u}^{\top} \right)\right]+\sigma \kappa \mathbf{n}\delta_s,
\end{gather}
where $\mathbf{u}$, $p$, $\kappa$, $\mathbf{n}$, and $\delta_s$ denote the velocity vector, pressure, interface curvature, interface unit normal vector, and Dirac delta function localized at the interface, respectively. The gas--liquid interface is captured using a VoF formulation, in which the interface is represented by the volume-fraction field $\chi$. Here, $\chi=1$ in the liquid phase and $\chi=0$ in the gas phase. The volume fraction is governed by a pure advection equation,
\begin{equation}
\partial_t \chi+\mathbf{u}\cdot\nabla\chi
=0.
\end{equation}

The numerical implementation of the Navier--Stokes equations in Basilisk closely follows that of Gerris \citep{popinet2003gerris,popinet2009accurate,lagree2011granular}, employing a projection method with second-order time integration, a CFL-limited timestep, the Bell--Collela--Glaz advection scheme, and an implicit viscosity solver. The governing equations are solved in a Cartesian coordinate system, $(x,y,z)$, where the $x$--, $y$--, and $z$--axes denote the coalescence direction, the wall-normal direction, and the direction orthogonal to both, respectively. The corresponding velocity components are $(u,v,w)$. Spatial discretization is performed on an adaptive octree (quadtree in 2D) Cartesian mesh \citep{popinet2015quadtree,van2018towards}. The mesh size and overall refinement strategy are selected based on the grid-convergence analysis presented in appendix~\ref{mesh}. The minimum grid size $\Delta_\text{min}/R_m$ is equal to 1/171 for the larger bubble and 1/85 for the smaller one. Grid adaptation is driven by wavelet-based error estimation for the velocity components, volume-fraction field, and concentration field. The refinement thresholds \citep{van2018towards} are set to $10^{-3}$ for the velocity field, $10^{-4}$ for the volume fraction field, and $10^{-1}$ for the concentration field. A geometric VoF method based on piecewise linear interface calculation (PLIC) \citep{scardovelli1999direct,tryggvason2011direct} is used for interface reconstruction and flux calculation of $\chi$. The interface curvature and interface normal vector are calculated using the well-balanced height-function method described by \cite{popinet:hal-01528255}.

For transport of species, we use the extension of the solver developed and validated by \cite{farsoiya2021bubble}, which neglects the variation in gas volume due to phase change. While the bubbles in real-world applications can grow, the time scale of this change is far larger than the time scale of coalescence; hence, the actual change of volume during the simulation window is a minor effect and does not justify the added numerical complexity. Specifically, as will be shown later, the bubbles evolve with an inertio-capillary time scale, $t_{ic}$. The ratio of $t_{ic}$ to the mass diffusion time scale, $t_d=R_m^2/\mathscr{D}$, is proportional to $\text{Oh}\cdot \text{Sc}^{-1}$ and ranges from $O(10^{-4})$ to $O(10^{-2})$ in the present study owing to the large Schmidt numbers and the small Ohnesorge numbers. The solver employs a single-field formulation \citep{haroun2010volume} in which the transport equation is solved for a single concentration field defined as $C=\chi C_l+(1-\chi) C_g$ over the entire domain. The transport of the variable $C$ is governed by
\begin{equation}
\partial_t C+\nabla\cdot(\mathbf{u}C)=\nabla\cdot\left[\mathscr{D}\nabla C-\mathscr{D}C\frac{\alpha-1}{\alpha \chi + (1-\chi)}\nabla \chi\right],
\end{equation}
where the single-field diffusion coefficient $\mathscr{D}$ is the harmonic mean of the diffusion coefficients of liquid and gas phases \citep{haroun2010volume},
\begin{equation}
\mathscr{D}=\frac{\mathscr{D}_l\mathscr{D}_g}{\mathscr{D}_g \chi + \mathscr{D}_l(1-\chi)}.
\end{equation}
The present formulation enforces the concentration jump associated with Henry's law, $C_l = \alpha C_g$, and the continuity of mass flux at the interface \citep{farsoiya2021bubble}. The solubility constant $\alpha=k_\mathrm{H}RT$ from Henry's law is set to 1/50, corresponding roughly to the properties of hydrogen and water at atmospheric pressure.  Note that properties other than diffusivity are calculated based on the arithmetic mean of the corresponding liquid and gas properties
\begin{equation}
\rho = \chi \rho_l + (1-\chi)\rho_g,
\qquad
\mu = \chi \mu_l + (1-\chi)\mu_g.
\end{equation}

\section{Results and Discussion}
\label{sec:results}
\subsection{Kinematics}
\label{sec:kinematics}
\begin{figure}[H]
\centering
  {\includegraphics[width=0.725\textwidth]{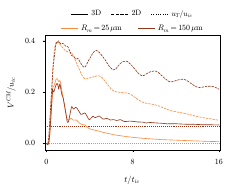}}
  
  \caption{Time evolution of the normalized center-of-mass vertical velocity for the two bubble sizes. Horizontal dotted lines indicate the normalized terminal velocity of an isolated rising bubble, with $u_T/u_{ic}=0.00131$ for $R_m=25\,\mu\mathrm{m}$ and $u_T/u_{ic}=0.06855$ for $R_m=150\,\mu\mathrm{m}$.}
\label{fig:vcm}
\end{figure}

Figure \ref{fig:vcm} shows the velocity of the center of mass of the bubbles after the moment of coalescence as a function of time. The velocity and time are both normalized in inertio-capillary units. For 3D bubbles, an initial rapid increase in velocity is observed as a result of the surface potential energy being converted to kinetic energy. The normalized velocities at two bubble sizes collapse well up to \textbf{$t/t_{ic}\approx 2$}, indicating that the velocity in this early stage of bubble rise scales well with the inertio-capillary velocity. A peak velocity at 0.23-0.25$u_{ic}$ is obtained for the 3D simulations; the peak velocity is slightly larger for the smaller bubble, which is the only aspect where the inertio-capillary velocity scaling is not perfect. A similar phenomenon has been reported for the jump-off of droplets on super-hydrophobic surfaces \citep{liu2014numerical}. Following the initial stage, the rise velocities of the larger and the smaller bubble clearly diverge. Setting aside some minor oscillations, both bubbles seem to asymptotically approach their `terminal' velocity at this stage. Evidently, the rise velocity does not scale with $u_{ic}$. If the bubble rise follows the Stokes law (which is a good approximation for at least the smaller bubble), the rise velocity normalized by $u_{ic}$ is of the order of $u_{\text{Stokes}}/u_{ic} \sim (\rho_l gR_m^2/\mu)/(\sigma/\rho_lR_m)^{1/2}\sim \text{Bo}/\text{Oh}$, which leads to two orders of magnitude difference between the two bubble sizes studied here. For reference, we add to the plot the values of terminal velocity $u_T$ for the 3D bubbles obtained from the balance of buoyancy force $4\pi g(\rho_l-\rho_g)R_m^3/3$ and drag force $\pi\rho_lc_
DR_m^2u_T^2/2$ on a spherical bubble using the drag coefficient $c_D$ from the work of \cite{mei1994note}.

\begin{figure}[H]
\centering
  {\includegraphics[width=1\textwidth]{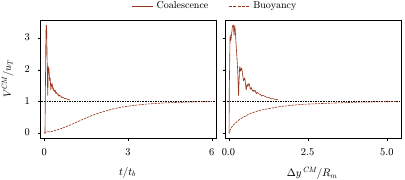}}
  
    \caption{Comparison of coalescing and isolated bubbles for $R_m=150~\mu\mathrm{m}$: normalized center-of-mass velocity versus time (left) and displacement, $\Delta y^{CM}/R_m$ (right).
}
\label{fig:velocity_displacement}
\end{figure}

The relatively abrupt shift away from the inerto-capillary scaling at $t/t_{ic}\approx 2$ is an interesting observation in figure \ref{fig:vcm}, which can be attributed to the near-complete dissipation of the released surface energy during this period \citep{cattani2026study}, after which the buoyancy and drag forces become the dominant forces. In the subsequent period, the bubbles decelerate asymptotically towards the corresponding terminal velocities, which, as described above, are significantly different for the two bubble sizes. Essentially, what we observe is the coalescence being followed by an immediate high-velocity jump-off lasting for a very short time ($\sim t_{ic}$), after which the free-rising terminal velocity is nearly recovered. At this point, it is insightful to compare the rise velocity after a coalescence to that of a spherical bubble with the same radius $R_m$ rising due to buoyancy from an initial position adjacent to the wall. Such a comparison is presented in the left panel of figure \ref{fig:velocity_displacement}, where both velocities are normalized by the terminal velocity of a bubble with $R_m=150\mu$m. While both coalescence-driven and buoyancy-driven bubble detachment eventually lead to the same terminal velocity, in the former case the initial velocity is much larger, but also the convergence to the terminal state is much faster. In the latter case, the asymptotic convergence has an expected time constant of approximately $t_b$. Note that the ratio $t_b/t_{ic}\sim \text{Bo}^{-1/2}$, which means that the separation in time scales grows for smaller bubbles. In the right panel of figure \ref{fig:velocity_displacement}, we plot the normalized rise velocities of the coalescence-driven and buoyancy-driven bubbles against the center-of-mass displacement, and it is clear that the rise velocities converge to the terminal value over a comparable vertical displacement. After a vertical displacement of $1.5R_m$ (the maximum available in the 3D coalescence-driven simulations), the $V^{CM}$ of the buoyancy-driven bubble is only about 22\% smaller than that of the coalescence-driven bubble.

\begin{figure}[H]
\centering

\begin{subfigure}{1\textwidth}
    \raggedright
    $(a)$\\
    \centering
    \includegraphics[width=1\textwidth]{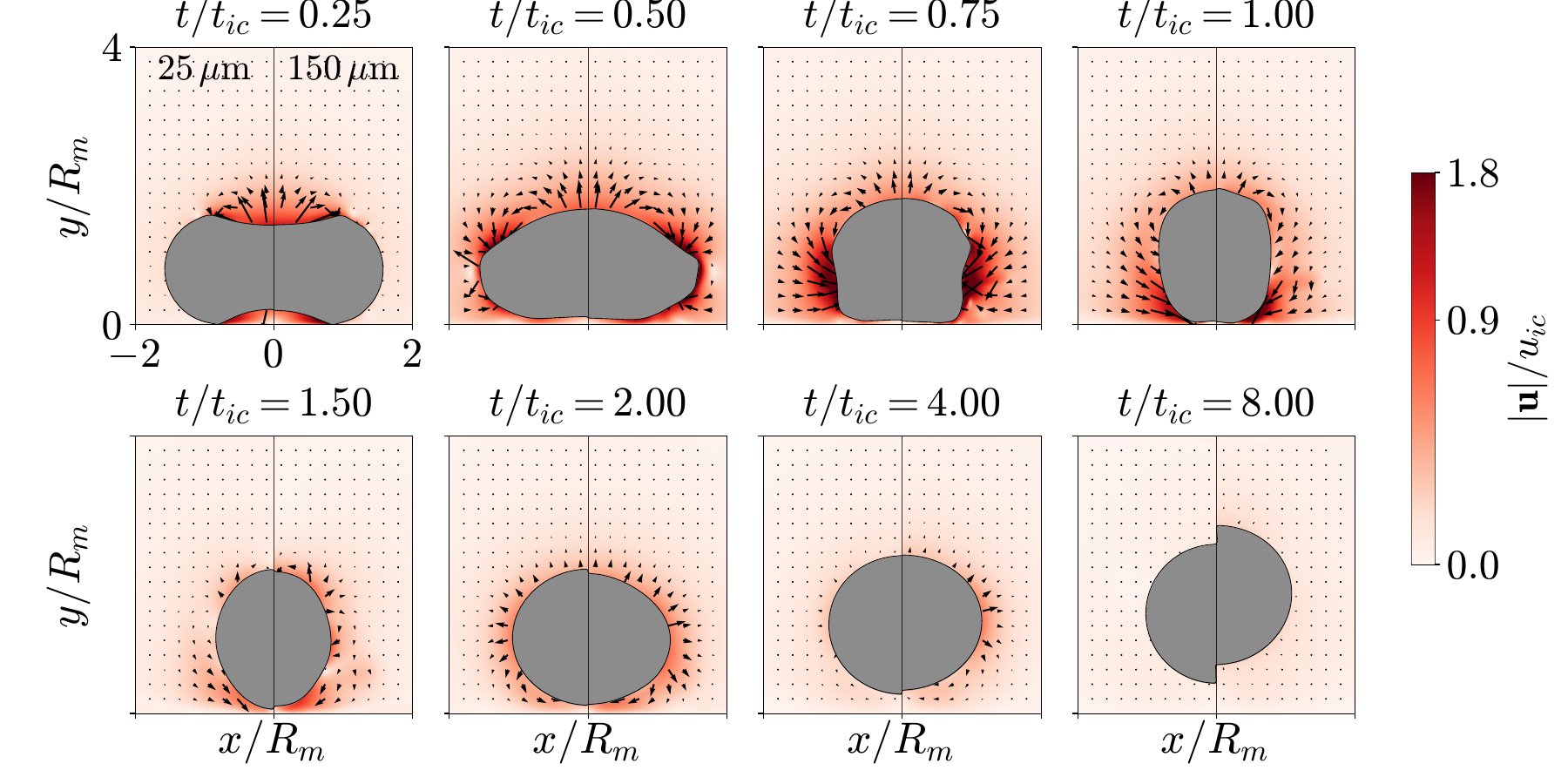}
\end{subfigure}

\vspace{0.5cm}

\begin{subfigure}{1\textwidth}
    \raggedright
    $(b)$\\
    \centering
    \includegraphics[width=1\textwidth]{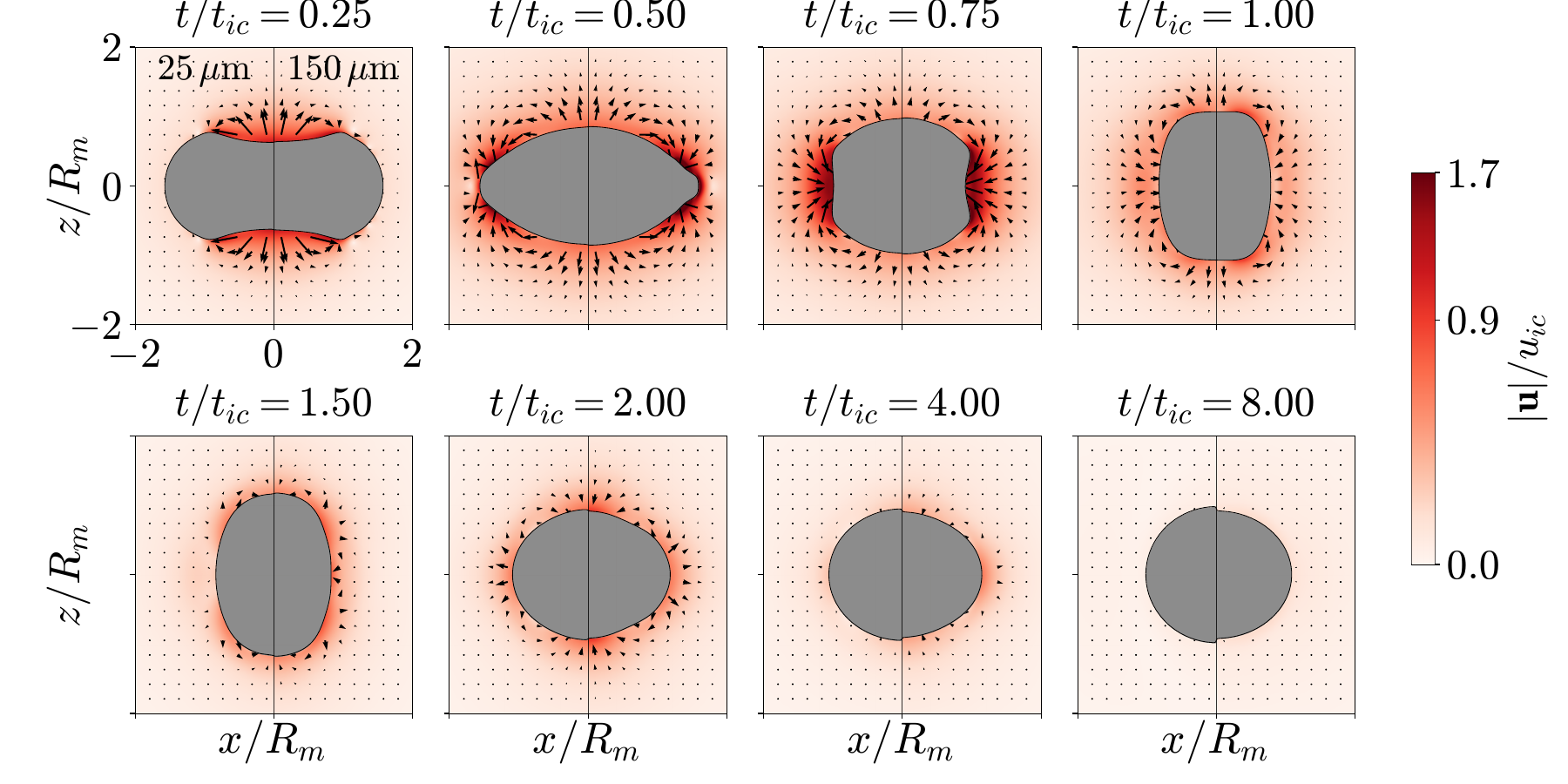}
\end{subfigure}
\caption{Instantaneous dimensionless velocity field around coalescing bubbles: (a) velocity fields in the symmetry $xy$-plane; (b) velocity fields in the $xz$-plane passing through the bubble's center of mass. Left and right halves correspond to $R_m=25\,\mu\mathrm{m}$ and $150\,\mu\mathrm{m}$, respectively. Colors denote the dimensionless velocity magnitude, $|\mathbf{u}|/u_{ic}$, and arrows indicate the local flow direction. The color range is different in (a) and (b).}
\label{fig:vel_contours}
\end{figure}

Figure \ref{fig:vel_contours} shows the bubble interface and the liquid velocity field around it for the two bubble radii at different normalized times after the coalescence. Both the side view (a) and the top view (b) indicate a rapid propagation of capillary waves from the point of first contact, which meet at the opposite apexes of the merged bubble at $t/t_{ic}\approx0.5$, forming a lemon-like shape. This is followed by an upward push due to the presence of the substrate, which eventually leads to the detachment of the bubble from the surface at a time between $t_{ic}$ and $1.5t_{ic}$. Similar to the center-of-mass velocity already discussed above, the overall interface velocity drops significantly at  $t/t_{ic}\approx2$. Generally, one observes the largest values of liquid velocity adjacent to the apexes around the moment of the lemon-like shape moving downwards as the time of jump-off approaches. Particularly, in the period $0.5\le t/t_{ic}\le 1$, a relatively strong downward velocity is observed around the lower side of the bubble, which, as will be discussed in the following, can meaningfully impact the transport. The shape of the interface is only slightly different for the smaller (left half) and larger (right half) bubbles up to the larger times when, as discussed before, the inertio-capillary scaling does not hold anymore, and bubbles rise with different velocities, hence the smaller bubble lagging behind. It is also observed in the top views that shortly after the jump off (approximately $t/t_{ic}>4$), the axial symmetry is nearly recovered and the bubbles rise in a near-spherical shape.

\begin{figure}[H]
\centering
  {\includegraphics[width=1\textwidth]{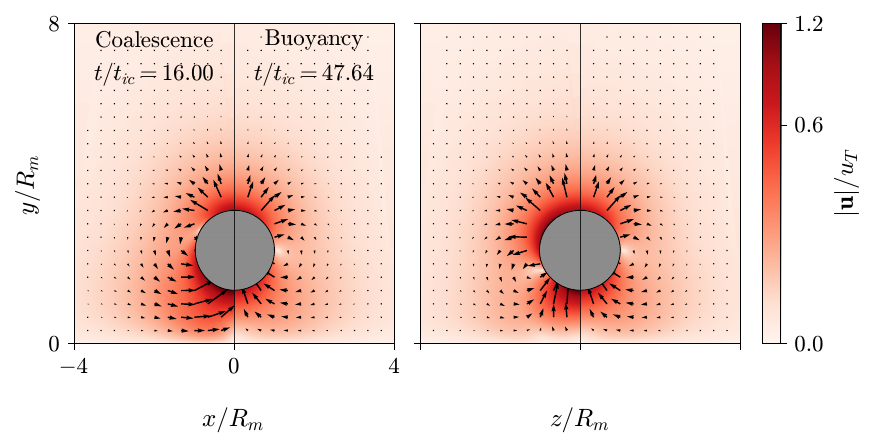}}
  
  \caption{Instantaneous velocity field normalized by the terminal velocity around coalescing bubbles (left halves) and a buoyancy-driven bubble (right halves). Left panel: velocity fields in the symmetry $xy$-plane; right panel: velocity fields in the $yz$-plane passing through the bubble's center of mass. Colors denote the normalized velocity magnitude, $|\mathbf{u}|/u_{T}$, and arrows indicate the local flow direction. The buoyancy-driven case is the result of an axisymmetric simulation. Bubble size is $R_m=150$ $\mu$m in both cases. }
\label{fig:axi_3d_vector}
\end{figure}

We depict the free rise of the bubble at larger times in figure \ref{fig:axi_3d_vector}, where, for better clarity, the velocity is normalized with the terminal velocity instead of $u_{ic}$. Here, we show two side views and provide a comparison to a purely buoyancy-driven rising bubble at the same center-of-mass location. As explained before, the second bubble, which starts from a spherical shape, takes longer to reach the same vertical location, as reflected in the time labels in the figure. Generally, the flow fields are relatively similar, both featuring a circular liquid motion, which extends down to near the wall. While the downwash due to this circular motion can play a role in mass transfer, one should note that the velocities in \ref{fig:axi_3d_vector} are far smaller than the early stages of bubble coalescence (figure \ref{fig:vel_contours}) as $u_b/u_{ic}\sim \text{Bo}^{1/2}$. For the coalescence-driven case, the $xy$- and $zy$-planes, while similar, are not identical, which indicates the motion at this point is not fully axisymmetric and holds certain memory of the initial coalescence phase.

\begin{figure}[H]
\centering
  {\includegraphics[width=1\textwidth]{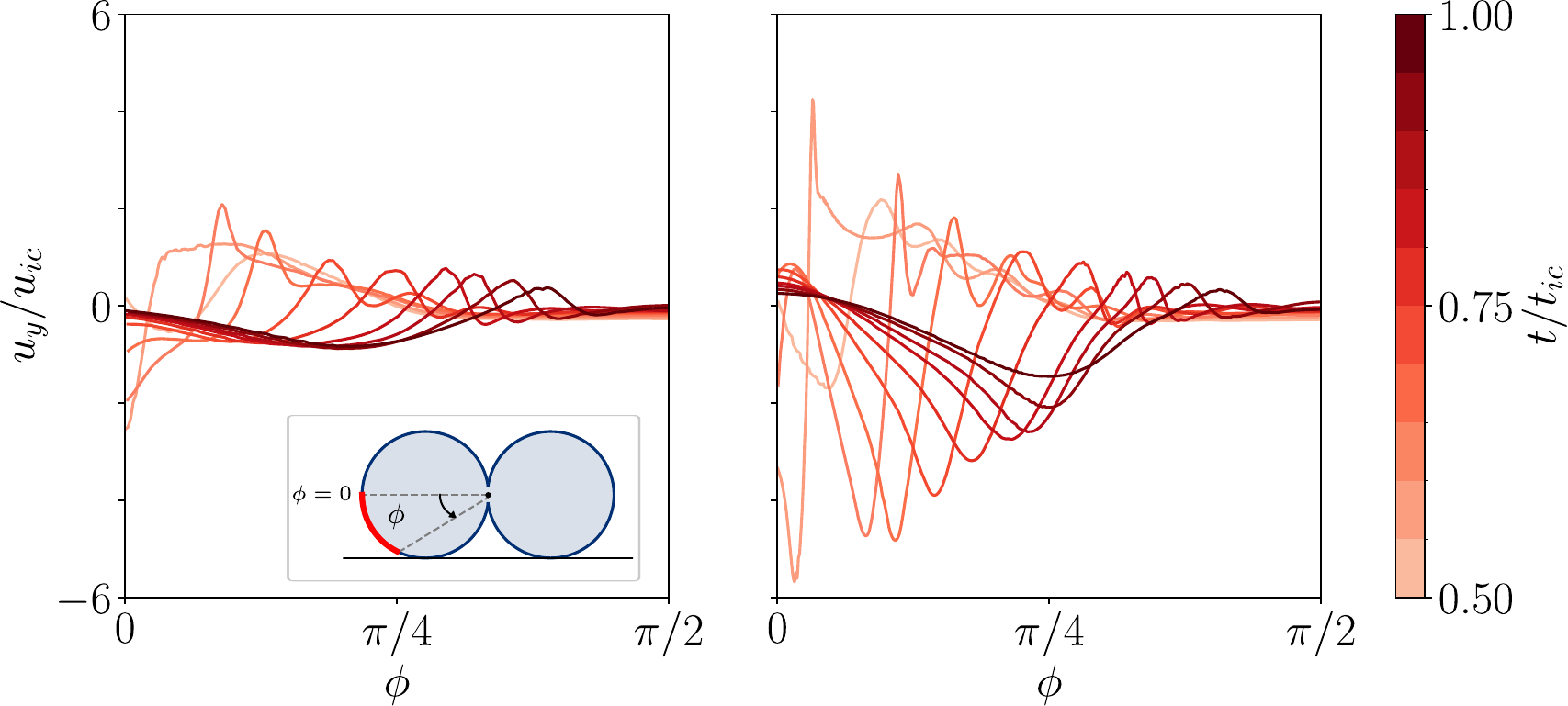}}
  \caption {Interface vertical velocity on the $z=0$ plane at varying the angular locations $\phi$, with reference to the initial center location of the parent bubble. Left is $R_m=25\ \mu\mathrm{m}$ and right is $150\,\mu\mathrm{m}$.}
\label{fig:velocity_angle}
\end{figure}

Since the bubble interface movement is the driving force for the liquid motion and, hence, for micro-convection, it is insightful to take a closer look at the interface velocity, particularly during the period when the largest velocities occur in the system. In figure \ref{fig:velocity_angle}, the interface velocities at the symmetry plane are depicted for both bubbles in the period $0.5\le t/t_{ic}\le 1$. We only plot the $y$-component of velocity, which is directly linked to the convection normal to the wall. Furthermore, we focus on the lower half of the bubble where the concentration gradient is pronounced. The plotted interface velocity curves clearly show the propagation of capillary waves towards larger angles over time. In both cases, a negative velocity peak is followed by a positive one before converging to zero around the bottom of the bubble ($\phi=\pi/2$), which is in contact with the substrate. Interestingly, while the general trend is similar for both bubble radii, there is a clear difference in the magnitude of the velocity; here, the larger bubble shows a very strong downward velocity, which is somewhat weakened for the smaller bubble. This reminds us that, even though the center-of-mass velocity and interface shapes are generally similar across different bubble radii when scaled in inertio-capillary units, there are deviations in local quantities. In particular, the interfacial velocity at the capillary wave front seems to be sensitive to bubble size within the inertio-capillary regime, particularly around the moment of lemon-like shape, and more severe damping is observed for smaller bubbles, where the effect of viscosity is more pronounced (larger Ohnesorge numbers).

\subsection{Mass transfer}
\label{sec:mass}
\subsubsection{Effect of bubble size}
\begin{figure}[H]
\centering

\begin{subfigure}{1\textwidth}
    \raggedright
    $(a)$\\
    \centering
    \includegraphics[width=1\textwidth]{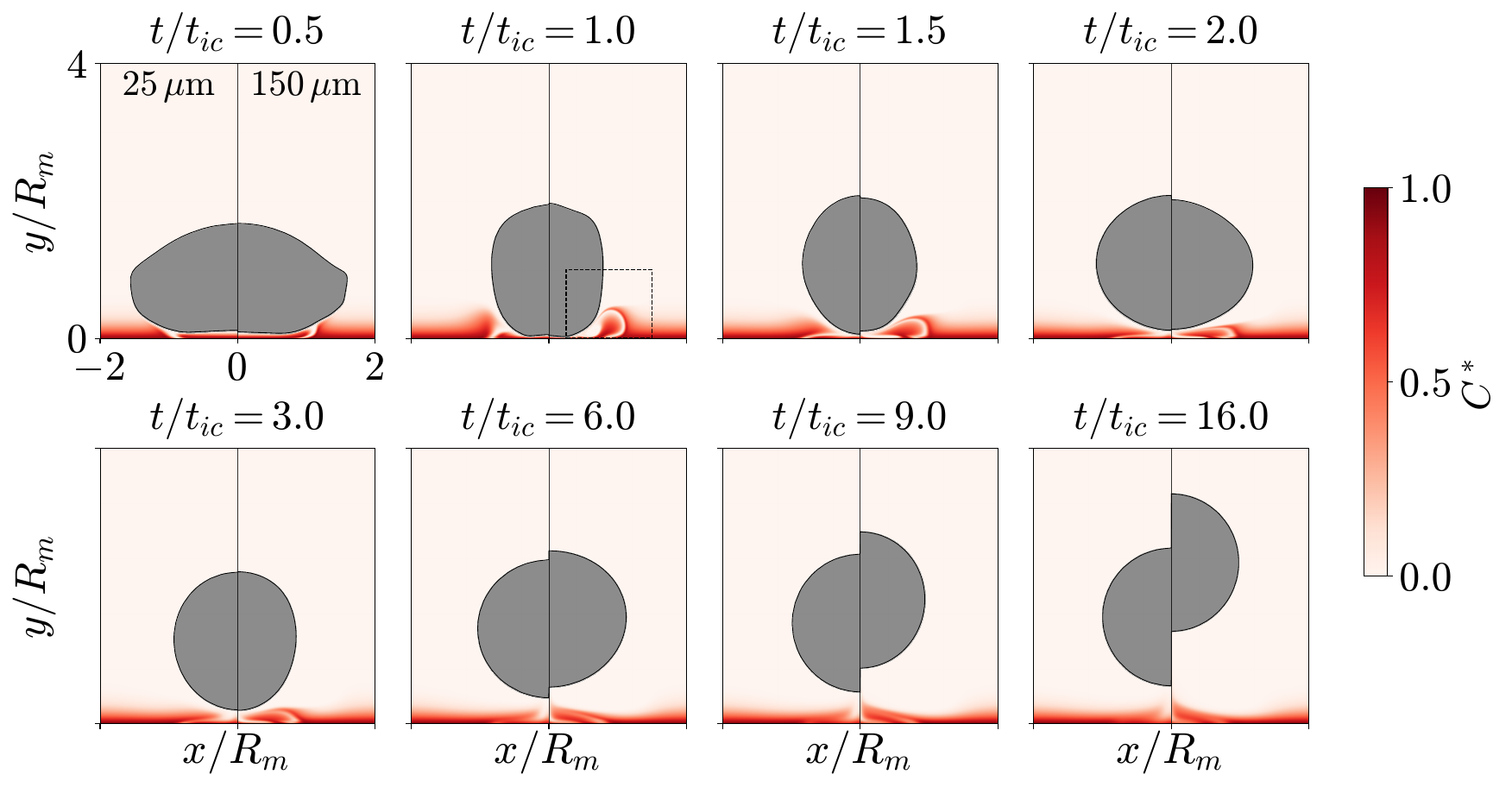}
\end{subfigure}

\vspace{0.8cm}

\begin{subfigure}{1\textwidth}
    \raggedright
    $(b)$\\
    \centering
    \includegraphics[width=1\textwidth]{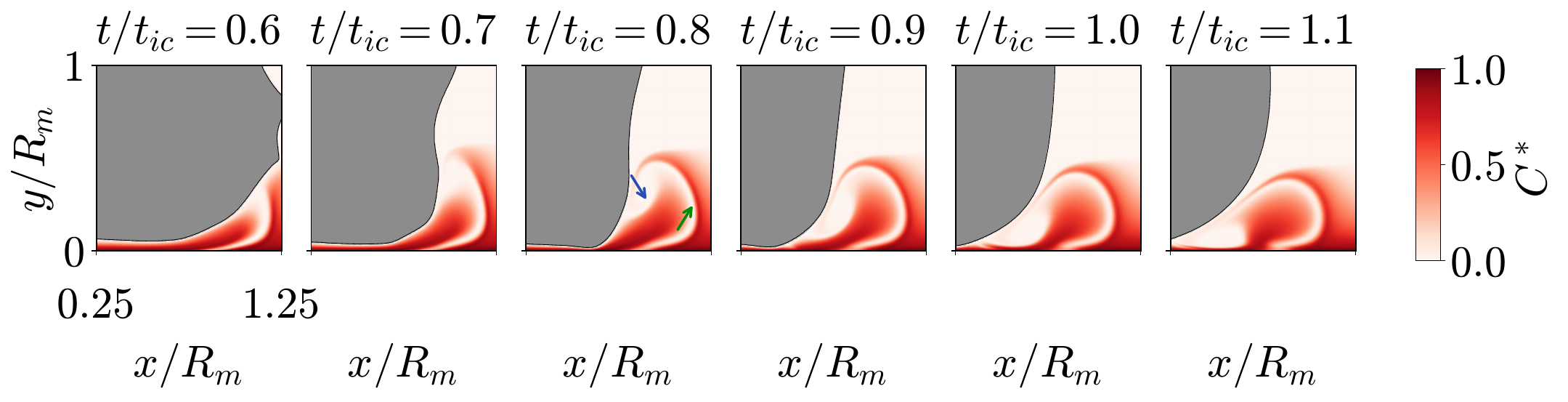}
\end{subfigure}
\caption{Normalized concentration field for $\mathrm{Sc}=210$. (a) Comparison of $C^*$ for the two bubble radii (left: $25~\mu$m; right: $150~\mu$m). (b) Magnified view of the boxed region in (a) for the $150~\mu$m case.}
\label{fig:D_contours}
\end{figure}

Figure \ref{fig:D_contours} shows the normalized concentration field $C^*$ on the $z=0$ plane for both bubble sizes. At this point, we consider only the cases with $\mathrm{Sc}=210$, corresponding to dissolved hydrogen in water given its direct application relevance. As a result of the high $\text{Sc}$ (low diffusivity), sharp concentration gradients are visible as the scalar transport approaches that of a passive tracer as $\text{Sc}\rightarrow\infty$. At $t/t_{ic}\approx1$, the concentration distribution develops a distinct pattern in the immediate vicinity of the bubble. Figure \ref{fig:D_contours}b presents a magnified view of this area, allowing a more detailed examination of the evolution of the concentration field. One can identify two low-concentration regions within the boundary layer, which are indicated by two arrows in the zoomed view. The first low-concentration region (green arrow) extends as a narrow trace away from the initial contact point of the parent bubble. In addition, between $t/t_{ic}=0.6$ and 0.7, a low-concentration packet of liquid (blue arrow) starts to appear adjacent to the bubble interface. This packet gradually moves towards the wall as it keeps deforming. Roughly at $t/t_{ic}\approx1.5$, this packet has evolved into a thin low-concentration layer next to the wall. As the process continues, the two low-concentration regions further deform, partly merge, and gradually fade out due to diffusion.

The first low-concentration region discussed above can clearly be traced back to the initial point of contact, where the proximity of the bubble interface and wall leads to a fast depletion of species concentration in the liquid. The formation of the second low-concentration region (the packet) coincides with the downward motion discussed under figure \ref{fig:velocity_angle}, and can be attributed to the entrainment of `fresh' liquid into the boundary layer. We furthermore observe that the distinction between the two regions is less evident at $R_m=25\mu$m compared to $R_m=150\mu$m. This can be due to at least two factors; first, the time scale of diffusion grows with the second power of length, here roughly $R_m^2$, while the time scale of interface movement ($t_{ic}$) grows with $R_m^{3/2}$. The ratio of the two time scales is therefore $R_m^{1/2}$, meaning that the diffusion acts relatively faster on the concentration field around the smaller bubble. Secondly, as indicated in figure \ref{fig:velocity_angle}, local interface movement is not identical for the two bubbles, which entails different patterns in the convection of the concentration field. Notably, the downwash is less intense at $R_m=25\mu$m, which creates a less pronounced low-concentration packet.

\begin{figure}[H]
\centering
  {\includegraphics[width=1\textwidth]{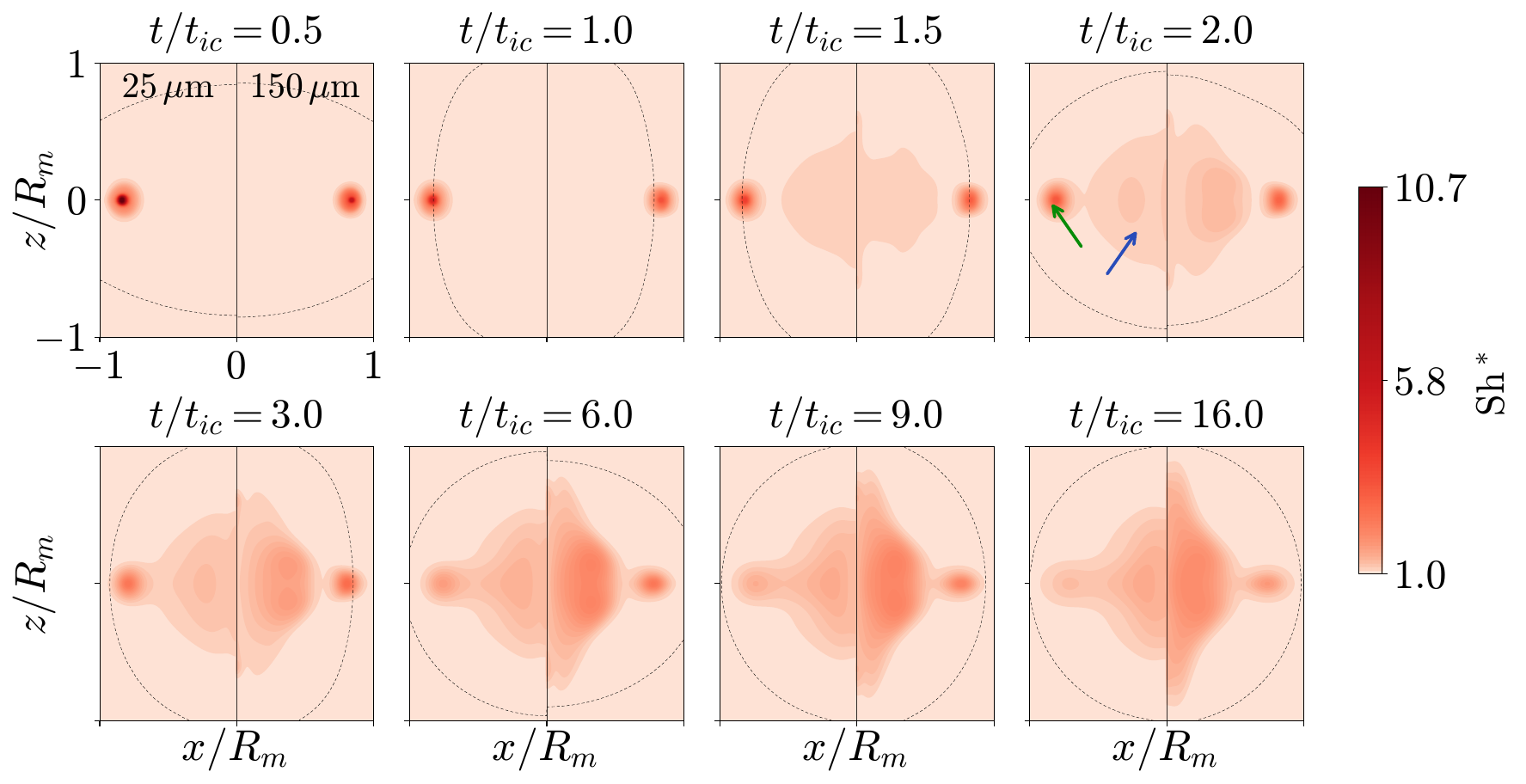}}
    \caption{Dimensionless Sherwood number at the wall on the $xz$-plane for $\mathrm{Sc}=210$. The left and right halves of each panel correspond to the $R_m=25\,\mu\mathrm{m}$ and $R_m=150\,\mu\mathrm{m}$ cases, respectively. Dashed curves indicate the projected bubble shadow. The green and blue arrows point at the `contact-point peak' and `enhanced-transfer core' described in the text, respectively.}
\label{fig:C_xz_Diamter}
\end{figure}

The effect of convection on wall mass transfer can be observed in  figure \ref{fig:C_xz_Diamter}, where the distribution of normalized Sherwood number $\text{Sh}^*$ on the wall is plotted for the two bubble sizes. As mentioned before, it should be noted that $\text{Sh}^*$ is normalized with the Sherwood number of pure diffusion at the corresponding time; hence, its value directly measures the mass transfer enhancement due to convection. For both bubble sizes, a very large, strong peak is observed in figure \ref{fig:C_xz_Diamter} at the location of initial contact -- contact-point peak hereafter.
The peak is deemed to be a result of the initial depletion of the dissolved species between the wall and bubble interface. As will be discussed in the following sub-sections, emergence of this peak is highly dependent on the initially prescribed concentration around the bubble. As the contact-point peaks gradually fade due to the effect of diffusion, an extended area with relatively high Sherwood number appears at $t/t_{ic}\approx1.5$ around the origin of the coordinates. This area -- enhanced-transfer core hereafter -- is well contained within a circle of smaller than $R_m$ radius around the center.
The emergence of the enhanced-transfer core can be attributed to the entrainment of the low-concentration packet, which has been discussed under figure \ref{fig:D_contours}. Note that at $t/t_{ic}\approx1.5$, this packet has morphed into a thick layer at the wall, and it is also around this moment that the bottom of the bubble starts moving upwards (see figure \ref{fig:vel_contours}). The enhanced-transfer core continues to increase in both size and intensity up to a certain point ($t/t_{ic}\approx10$). Due to the high computational cost, we do not continue the simulation beyond $t/t_{ic}=16$, but further smoothing is expected at larger times. While the overall pattern is similar for both bubbles, the smaller bubble shows smaller values of $\text{Sh}^*$, which is a reflection of the less pronounced entrainment of liquid at the early stages of bubble lifetime as discussed in section \ref{sec:kinematics}. Finally, the shape of the enhanced-transfer core exhibits two symmetry axes, as expected, but it is clearly not axisymmetric despite the fact that both the bubble shape and the induced velocity field become nearly axisymmetric towards the end of the simulation. Indeed, the mass transfer coefficient remains strongly dependent on the history of bubble motion and its effect on the concentration field. This history effect persists particularly because of the slow diffusion at large Schmidt numbers.

\subsubsection{Comparison of 3D and 2D problems}
To limit the number of costly 3D simulations, the effects of certain parameters are studied using 2D simulations in the present work. It is, therefore, necessary to explore the similarities and differences in the physics of mass transfer in these two configurations. Furthermore, such a comparison can shed light on the contribution of the $z$-component of velocity in the observed mass transfer enhancement.

\begin{figure}[H]
\centering
  {\includegraphics[width=01\textwidth]{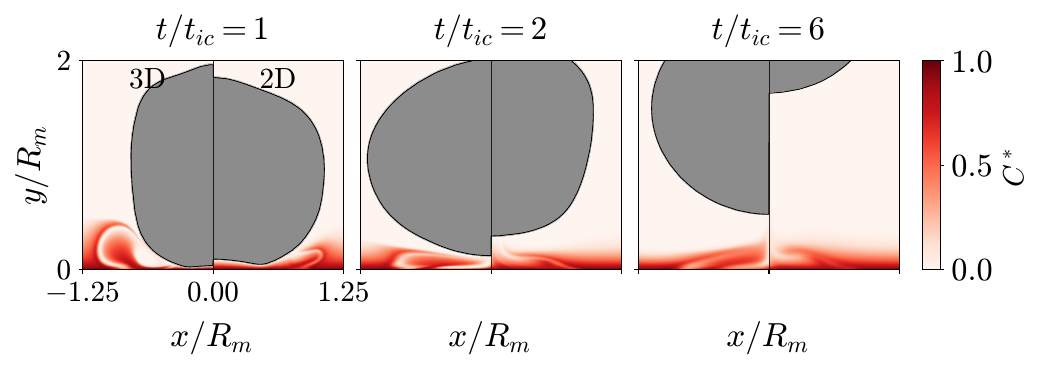}}
  \caption{Comparison of the normalized concentration field in three-dimensional (left half of each panel) and two-dimensional (right half of each panel) simulations for $R_m=150\,\mu\mathrm{m}$ at $\mathrm{Sc}=210$.}
\label{fig:2d3d_c}
\end{figure}

Figure \ref{fig:2d3d_c} provides a side-by-side comparison of the bubble interface shapes and concentration fields obtained from the 2D and 3D simulations. In both simulations, the radius of the merged bubble ($R_m=$150 $\mu$m) and the initial concentration distribution ($\delta_i=R_m/3$) are kept identical. Although the bubble evolution in the two cases represents similar overall patterns, characterized by an initial contraction followed by a subsequent jump-off event, noticeable differences arise in the exact interface shape and resulting jump velocity. Such differences are only expected as the 2D bubble lacks the second principal curvature, which clearly influences the dynamics.
Evidently, different interface dynamics translate into different surrounding flow fields, which is eventually reflected in the liquid entrainment patterns during the pre-jump phase, and consequently, in a smaller low-concentration packet forming farther away from the wall in the 2D case.

\begin{figure}[H]
\centering
  {\includegraphics[width=1\textwidth]{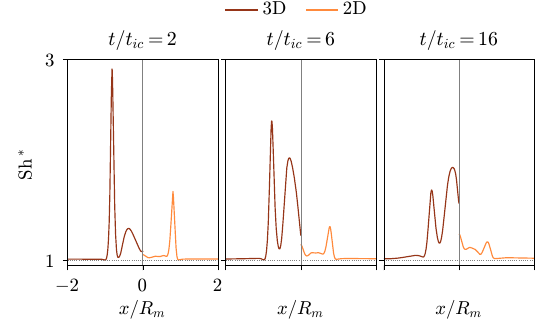}}
  \caption{Normalized Sherwood number from 3D (left halves) and 2D (right halves) simulations for $R_m=150~\mu\mathrm{m}$ and $\mathrm{Sc}=210$.}
\label{fig:2d3d_sh}
\end{figure}

Distributions of normalized Sherwood number resulting from the 2D and 3D concentration fields are, furthermore, displayed in figure \ref{fig:2d3d_sh}. For the 3D case, the computed values of $\text{Sh}^*$ are plotted along the $x-$axis. We observe that the contact-point peak emerges prominently in the 2D simulation, similar to the 3D case, representing a sharp local increase in $\text{Sh}^*$, which continuously fades as time goes by. The 2D case also exhibits a region of elevated $\text{Sh}^*$ between the contact-point peak and the center. While the Sherwood number in this region is noticeably smaller than that observed in its 3D counterpart, the region can nevertheless be considered analogous to the intense-transfer core discussed previously. The quantitative differences between the two configurations can be attributed to the distinct interface velocity patterns, which in turn lead to differences in both the shape and wall distance of the entrained low-concentration packets. Overall, although the 2D and 3D cases differ considerably in terms of the computed Sherwood number values and the detailed convection patterns, they exhibit a similar qualitative feature: a localized region of elevated mass transfer around the center as a result of micro-convection induced by bubble interface motion prior to the jump.

\subsubsection{Effect of Schmidt number}
\begin{figure}[H]
\centering
  {\includegraphics[width=1\textwidth]{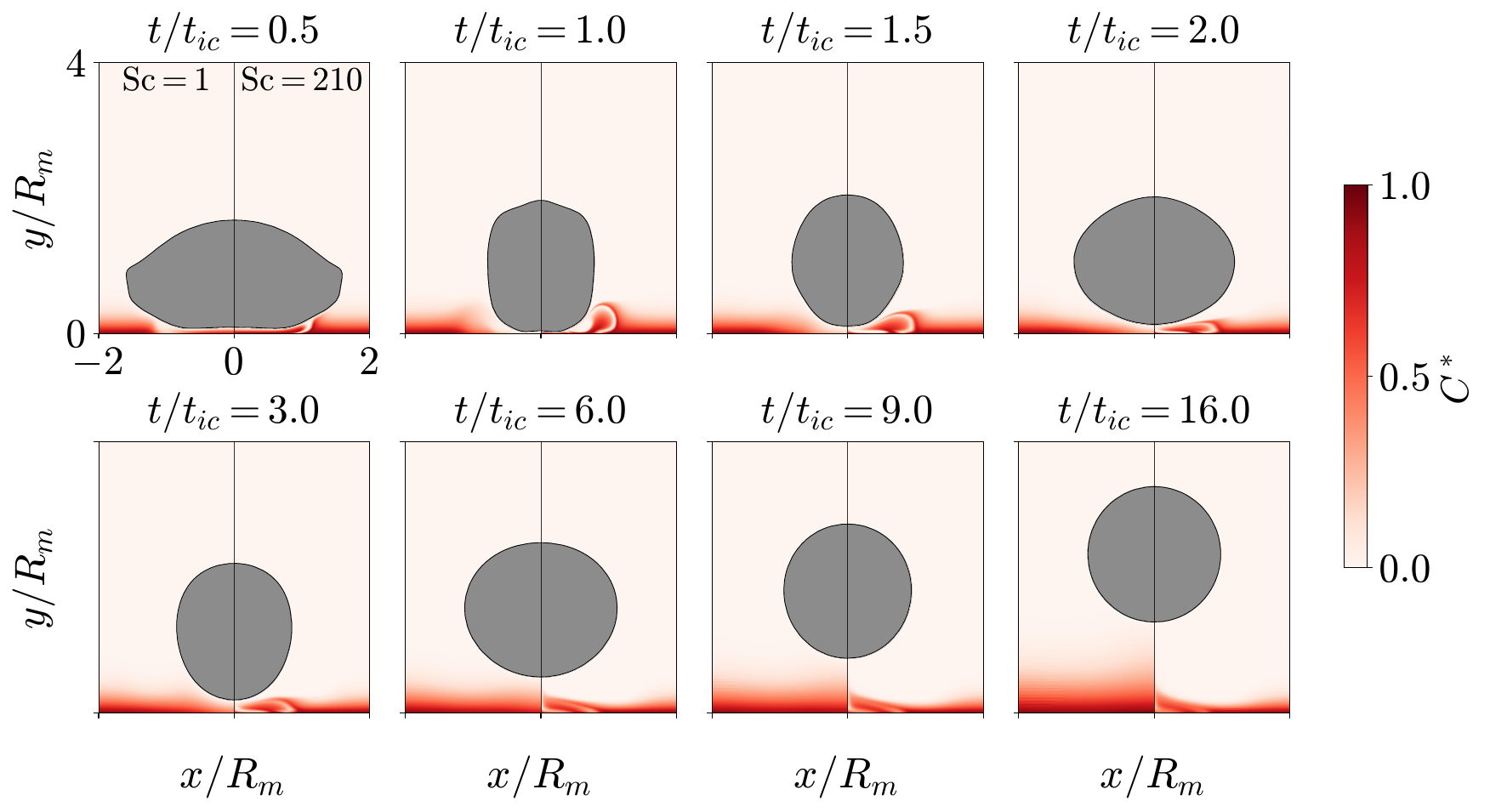}}
  \caption{Normalized concentration fields for $R_m=150~\mu\mathrm{m}$. The left and right halves show the $\mathrm{Sc}=1$ and $\mathrm{Sc}=210$ cases, respectively.}
\label{fig:sc_comp}
\end{figure}

Figure \ref{fig:sc_comp} depicts the concentration fields for a bubble of $R_m=150$ $\mu$m at two different values of Schmidt number, $\text{Sc}=1$ and 210. Evidently, fast diffusion at  $\mathrm{Sc}=1$ entails a dramatic change in the concentration field near the bubble, with sharp gradients completely smeared out due to the effect of diffusion. Indeed, unlike the case with $\mathrm{Sc}=210$, no clear distinction between the low-concentration regions due to the depletion and the wall-normal entrainment effects is observed. It is instructive to consider the length $\ell$, upon which diffusion acts in one inertio-capillary time unit (the time scale of convection in the present problem). Simply, equating the diffusion time scale $\ell^2/\mathscr{D}_l$ with $t_{ic}$ yields $\ell/R_m=\sqrt{\text{Oh/Sc}}$. Using this estimate, the diffusion length $\ell$ is about 10\% of the bubble radius at $\mathrm{Sc}=1$ compared to less than 1\% at $\mathrm{Sc}=210$ (for the smaller bubble size -- not shown -- the values of $\ell/R_m$ are larger by a factor of about 1.6). Considering the typical size of the low-concentration features (see, e.g., the zoomed view in figure \ref{fig:D_contours}), this estimate of $\ell$ can clearly explain why such features persist at $\mathrm{Sc}=210$ but are smeared out at $\mathrm{Sc}=1$.

After the bubble has jumped off, in both cases, a thickening of the boundary layer starts to emerge below the south pole of the bubble at $t/t_{ic}\sim 10$. For larger Schmidt numbers, however, some low-concentration features remain embedded within the thickened boundary layer. As discussed above, these features sustain the mass transfer coefficient at an elevated level for a longer period. In other words, the slower diffusion causes the concentration boundary layer to retain a `memory' of the transient micro-convection long after the velocity field has decayed. One should note that, during the time window of simulation ($t/t_{ic}\le16$), the concentration layer on average grows faster at $\text{Sc}=1$, in line with the 1D analytical solution. Finally, after about 10 time units, a slight thinning of the boundary layer is also observed at around one radius off center, which can be attributed to the weak downward motion of liquid discussed under figure \ref{fig:axi_3d_vector}.

Figure \ref{fig:sc_comp} shows the variation of normalized Sherwood number for all studied combinations of $\text{Sc}$ and $R_m$ along the $x-$axis. At $\text{Sc}=1$, a less-pronounced contact-point peak is observed initially. This peak quickly transitions to the center as the jump-off moment approaches, which is also when the bubble's south pole touches the substrate. It is arguable that, while both depletion and entrainment effects are in action, no clear distinction between their wall mass-transfer footprints can be made. Once the bubble has jumped off, apart from minor liquid circulation, diffusion is the main transport mechanism, and $\text{Sh}^*$ is expected to approach a uniform distribution, as the decay of its peak confirms. At $\text{Sc}=1$ and for the larger bubble size, the last shown moment indicates formation of a small minimum at the center flanked by two small peaks. This can be linked to a thickened boundary layer that suppresses mass transfer in the middle. While the Sherwood number trend at $\text{Sc}=210$ has been discussed previously, the distinction between the contact-point peak and the core region makes a clear contrast with the cases at $\text{Sc}=1$. Also the difference in the mass transfer coefficient for the larger and smaller bubble can be more clearly observed in these plots. Quantitatively, the ratio of mean Sherwood numbers at the two sizes (averaged over an arbitrary area of $R_m/2$ radius around the center), is 1.3 during the period between $t/t_{ic}=3$ and 16. Given that the jump-off Reynolds number, defined with $u_{ic}$ and $R_m$, is $1/\text{Oh}$, the ratio of the Reynolds numbers at two sizes equals 2.44. This yields a $\text{Sh}\sim \text{Re}^{0.3}$. Obviously, the power-law approximation is largely speculative, as establishing the exact functional relation with only two data points is not feasible.

\begin{figure}[H]
\centering
  {\includegraphics[width=1\textwidth]{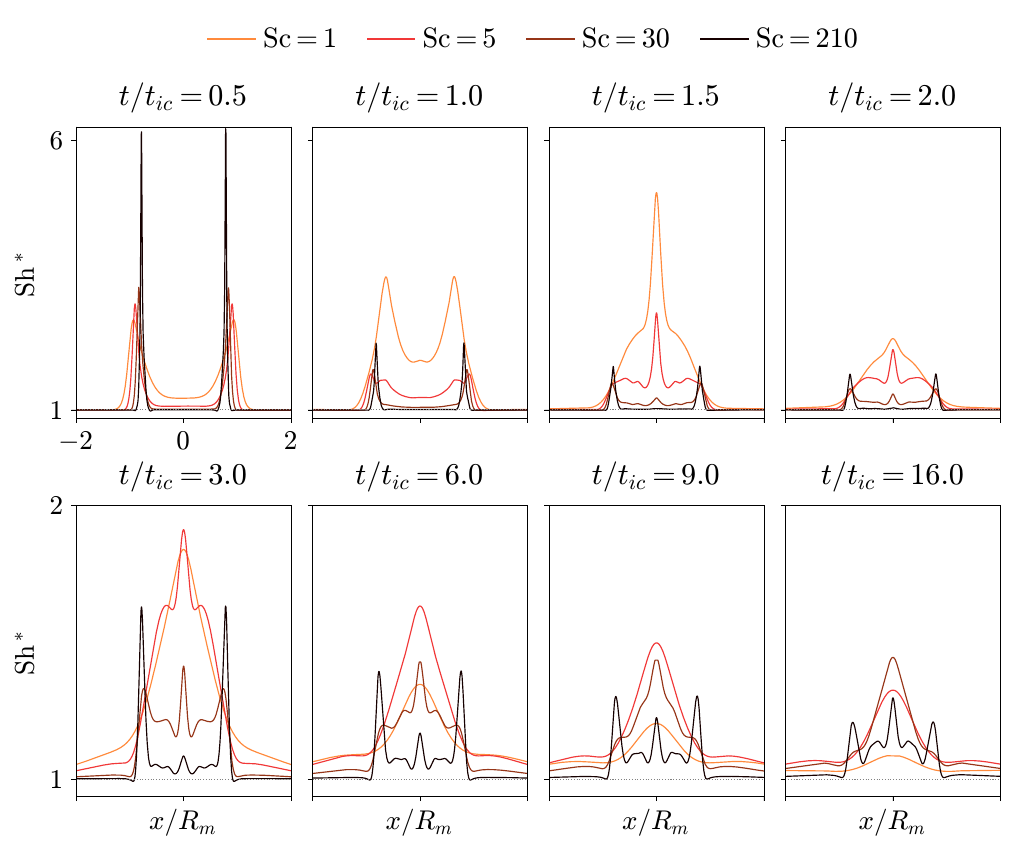}}
\caption{Normalized Sherwood number from 2D simulations for different values of $\mathrm{Sc}$. The bubble radius is $R_m=150\mu$m. Note the difference in axis scales in the two rows.}
\label{fig:scComp}
\end{figure}

To provide a more complete picture of the transition between the two values of $\text{Sc}$ discussed above, we conducted 2D simulations including two intermediate values of $\text{Sc}=5$ and 30. The results are plotted in figure \ref{fig:scComp}. As discussed before, the 2D problem involves fundamental differences with the 3D; nevertheless, certain features such as the presence of a core region with an elevated mass transfer coefficient are qualitatively similar. The results in figure \ref{fig:scComp} indicate a clear transition in terms of both mass transfer patterns and the diffusion time scale.


\begin{figure}[H]
\centering
  {\includegraphics[width=1\textwidth]{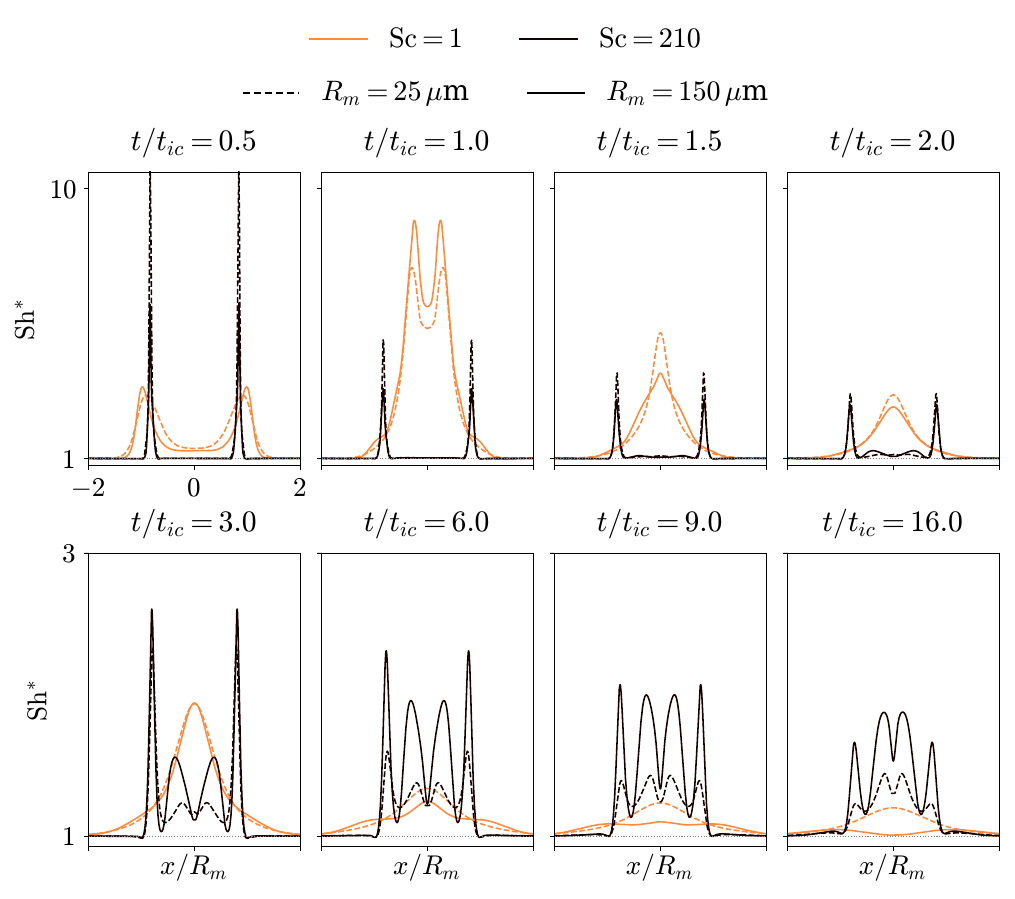}}
  \caption{Evolution of normalized Sherwood number from 3D simulations for two values of $\mathrm{Sc}$ and two bubble radii on the $z=0$ line. Note the difference in axis scales in the two rows. }
\label{fig:scComp_3D}
\end{figure}

\subsubsection{Effect of initial concentration boundary layer}
\label{sec:bl}
\begin{figure}[H]
\centering
  {\includegraphics[width=1\textwidth]{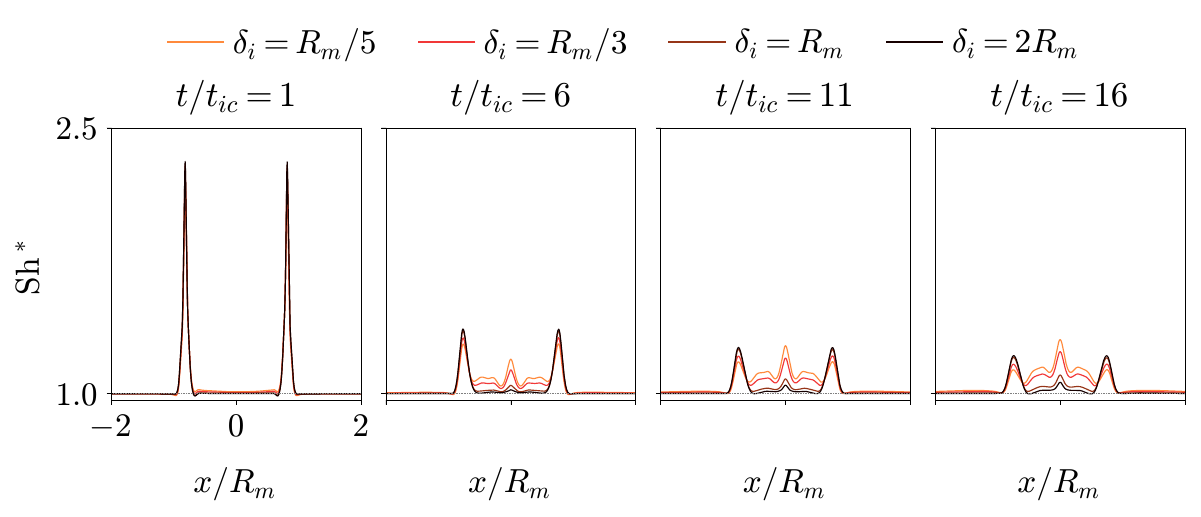}}
    \caption{Evolution of the normalized Sherwood number from 2D simulations at $\mathrm{Sc}=210$ for different initial concentration boundary-layer thicknesses. The bubble radius is $R_m = 150\,\mu\mathrm{m}$.}
\label{fig:deltaComp}
\end{figure}

To shed further light on the influence of the initial concentration boundary layer at the moment of coalescence, additional 2D simulations are conducted in the present work. Firstly, we vary the initial boundary layer thickness $\delta_i$, defined in section \ref{sec:problem}. The time evolution of the normalized Sherwood number at $\text{Sc}=210$ and $R_m=150$ $\mu$m is plotted for different values of $\delta_i$ in figure \ref{fig:deltaComp}. We observe the most pronounced influence of $\delta_i$ on $\text{Sh}^*$ to emerge in the core region. This aligns with the expected driving mechanism of mass transfer enhancement in this area: because the downward motion responsible for entraining the fresh liquid into the boundary layer acts within a specific height range, thicker boundary layers reduce its effectiveness. Conversely, the contact point peaks remain mostly unaffected by $\delta_i$ because their origin is independent of the convection; they only diverge at later times due to the diffusion effect.

\begin{figure}[H]
\centering
  {\includegraphics[width=1\textwidth]{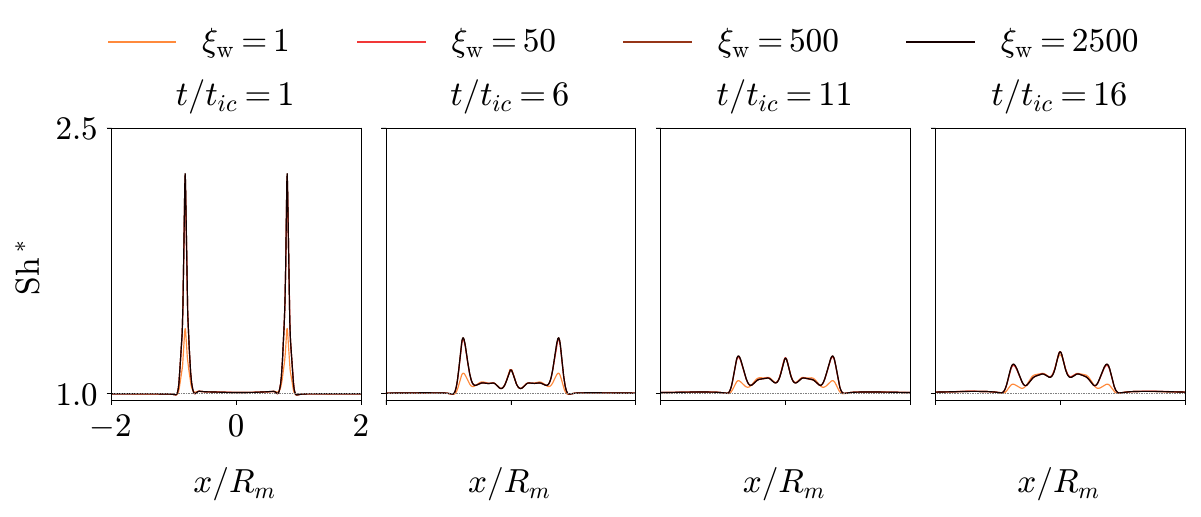}}
    \caption{Evolution of the normalized Sherwood number from 2D simulations at $\mathrm{Sc}=210$ for different values of initial wall supersaturation, $\xi_\mathrm{w}$. The bubble radius is $R_m = 150\,\mu\mathrm{m}$.}
\label{fig:ksiComp}
\end{figure}

\begin{figure}[H]
\centering
  {\includegraphics[width=1\textwidth]{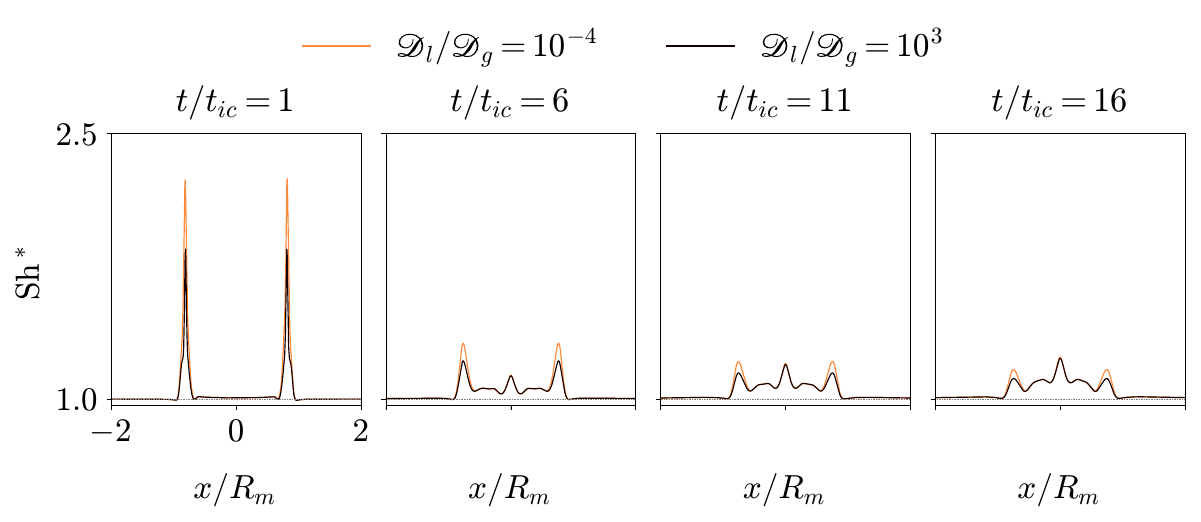}}
    \caption{Evolution of the normalized Sherwood number from 2D simulations at $\mathrm{Sc}=210$ for two different values of diffusion coefficient ratio ($\mathscr{D}_l/\mathscr{D}_g=10^{-4}$ and $10^{3}$). The bubble radius is $R_m = 150\,\mu\mathrm{m}$.}
\label{fig:DReffect}
\end{figure}

We furthermore investigate two parameters directly affecting the mass transfer across the bubble interface, namely the supersaturation level in figure \ref{fig:ksiComp}, and the diffusion coefficient of the gaseous phase in figure \ref{fig:DReffect}. It is observed in figure \ref{fig:ksiComp} that only the smallest considered value of supersaturation, $\xi_w=1$, leads to a different result in form of a smaller contact-point peak of $\text{Sh}^*$. This can be justified by the fact that a smaller $\xi_w$ translates to lower interfacial mass transfer into the bubble, which, as explained before, is deemed to be the reason for the existence of the peak. Note that, in real-world scenarios and at sufficiently large current densities, supersaturation is several orders of magnitude above one \cite[see, e.g.,][]{khalighi2023hydrogen,sepahi2024mass}. Indeed, following the above finding, we adopt a constant large value of $\xi_w=500$ across all simulations in the present work.

Additionally, we run simulations in which an unrealistically small relative diffusion coefficient is prescribed for the gas phase ($\mathcal{D}_g/\mathcal{D}_l=10^{-3}$). The reason for running this extreme numerical experiment is to gain insight into the sensitivity of the measured Sherwood numbers to interfacial mass transfer in the present setup. As expected, the results show that the added resistance on the gas side leads to a suppression of the contact point peak, but no significant change to $\text{Sh}
^*$ is observed elsewhere. Overall, both figures \ref{fig:ksiComp} and \ref{fig:DReffect} indicate that the mass transfer coefficient in the enhanced-transfer core region is virtually insensitive to the parameters affecting the interfacial mass transfer, and it is only the contact-point peak that can be affected by those. 

\subsection{Discussion}
\label{sec:discussion}
The primary question of interest in the present work is whether, and to what extent, micro-convection can enhance wall mass transfer in the wake of a detaching bubble. While the results in section \ref{sec:mass} indicate elevated values of wall Sherwood number following bubble coalescence, further discussion on the real-world implications of these observations is essential since, strictly speaking, the present generic problem setup is not identical to that of hydrogen and oxygen bubbles growing on an electrode surface.

\begin{figure}[H]
\centering
  {\includegraphics[width=0.55\textwidth]{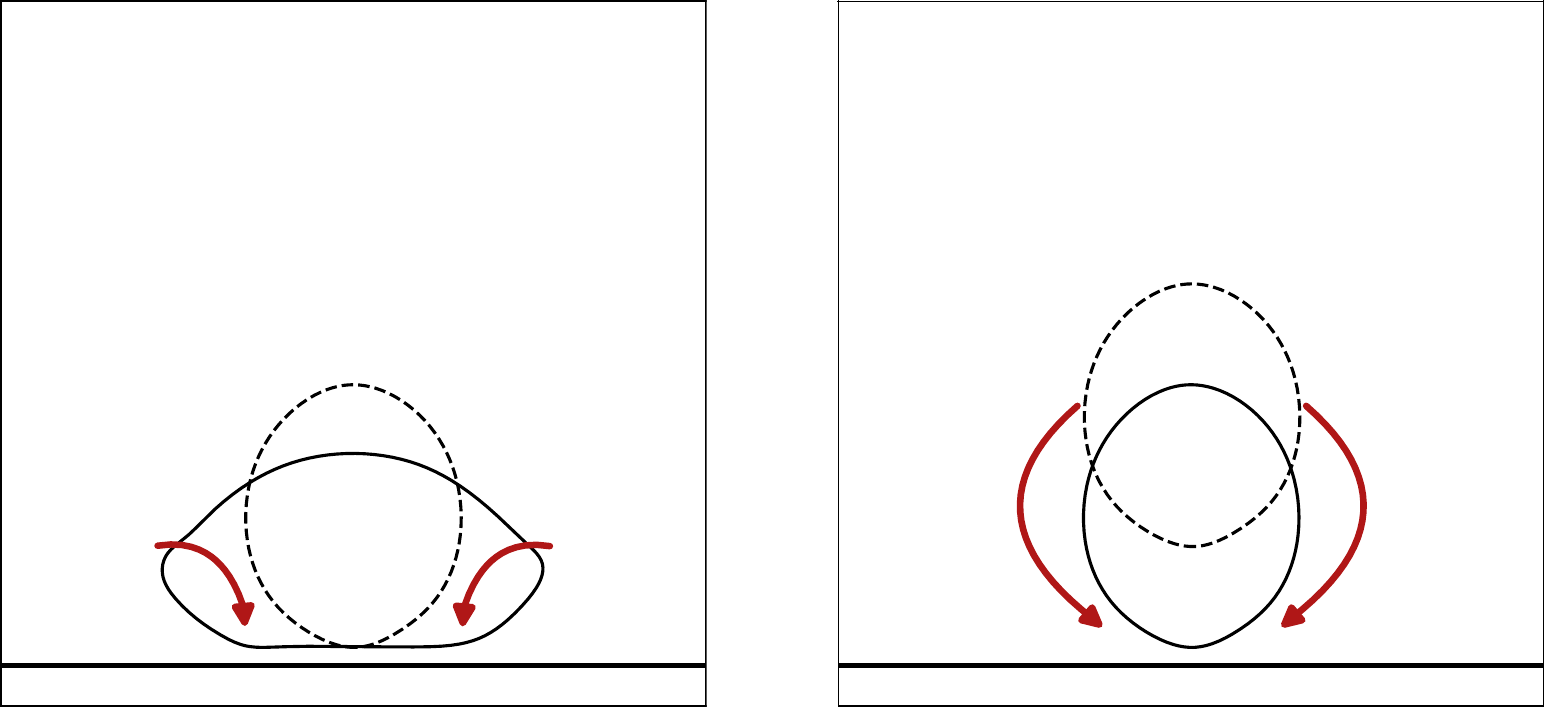}}
    \caption{Schematics of the micro-convection patterns (red arrows) induced by initial bubble deformation (left) and the following bubble rise (right).}
\label{fig:convection}
\end{figure}

According to the classical view \citep{vogt2015local,zhao2019gas}, bubble-induced convection is linked to either the growth of a bubble or its motion and the displacement of species-carrying liquid caused by them. In the present work, bubbles do not grow; therefore, only the latter micro-mechanism is in action. In section \ref{sec:kinematics}, we observed two distinct patterns in the liquid velocity that can impact wall mass transfer: at the early stages after the coalescence, when the inertio-capillary forces are dominant, pronounced liquid movements is observed close to the bubble interface, whose most noticeable manifestation is a downwash around the southern hemisphere of the parent bubble in the period $0.5<t/t_{ic}<1 $. Once the initial fast evolution of the bubble interface has decayed, the buoyancy-induced bubble rise creates a circulation of the surrounding liquid. These two flow patterns are schematically depicted in figure \ref{fig:convection}. It should be borne in mind that the velocity scales with $u_{ic}$ and $u_{b}$ in the first and second stages, respectively, meaning that, for sub-millimeter bubbles in water, the first mechanism is meaningfully stronger. The Reynolds number based on these two characteristic velocities is equal to $1/\text{Oh}$ and $\sqrt{\text{Bo}}/\text{Oh}$ -- the former being considerably larger. One should note that these two mechanisms both act at bubble scale and differ from the buoyancy-induced bulk mixing \citep{janssen1979effect,sepahi2024mass}; therefore, the term micro-convection is deemed appropriate for both.

We now specifically turn attention to the first mechanism mentioned above. Results in section \ref{sec:mass} show an increase in the normalized Sherwood number, $\text{Sh}^*$, in a region referred to as the intense-transfer core, which we attribute to the mechanism in question. Specifically, the liquid downwash linked to this mechanism leads to a visible disturbance of the concentration boundary layer and its local thinning. It must be mentioned that, in the present simulations, low concentrations can also be created locally inside the boundary layer and around the point of wall-contact of the parent bubbles, which is linked to the intense mass transfer due to proximity of bubble interface. The observed `contact-point' peak of $\text{Sh}^*$ is attributed to this effect. It is specifically shown in section \ref{sec:bl} that a variation in parameters such as initial supersaturation and mass diffusivity of the gas phase, which affect the interfacial mass transfer into the bubble, only modify the contact-point peak and not the enhanced-transfer core, which is in-line with the present explanation.

One must note that the phenomenon leading to the contact-point peak cannot be regarded as micro-convection and indeed it is primarily an artifact of the adopted constant flux wall boundary condition; in real-world, the electric current density and the resulting flux of species on an electrode are not uniform and the coupling between the electric potential, concentration and volume fraction fields suppresses the flux in the vicinity of the contact point. Therefore, making a distinction between the two effects is crucial. Consequently, in an attempt to isolate the enhancement in $\text{Sh}^*$ due only to the micro-convection, we run an additional simulation, in which the concentration boundary layer is `re-started' at time $t/t_{ic}=0.5$, which is the approximate start time of the downwash. At this time, we prescribe a 1D concentration gradient, which clears any patterns in the concentration boundary layer created due to the interactions prior to the start of the downwash.

\begin{figure}[H]
\centering
  {\includegraphics[width=0.8\textwidth]{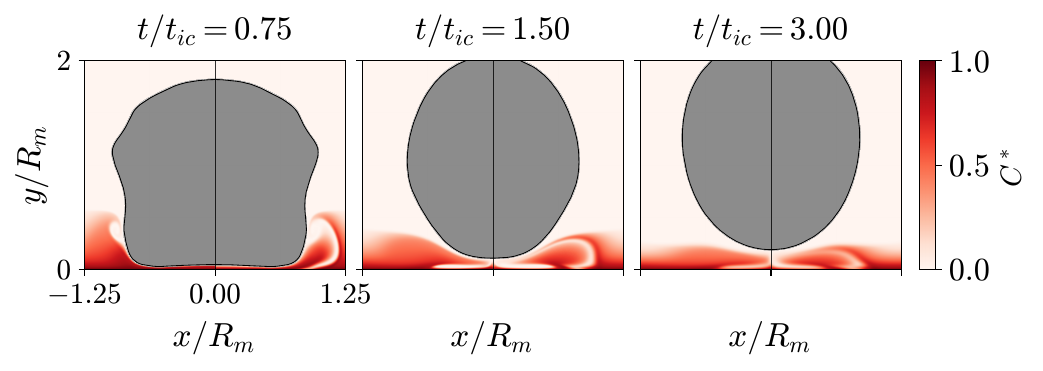}}
  \caption{The normalized concentration field obtained from the original simulation (right halves) and the simulation with the concentration field restarted at $t/t_{ic}=0.5$ (left halves). The simulations are conducted at $\text{Sc}=210$ and $R_m=210$ $\mu$m.}
\label{fig:restart_c05}
\end{figure}

\begin{figure}[H]
\centering
  {\includegraphics[width=0.5\textwidth]{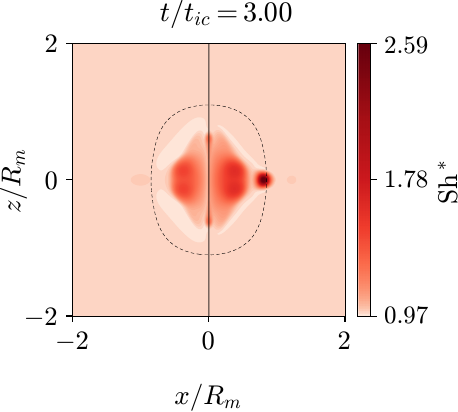}}
  \caption{The normalized wall Sherwood number obtained from the original simulation (right half) and the simulation with the concentration field restarted at $t/t_{ic}=0.5$ (left half). The simulations are conducted at $\text{Sc}=210$ and $R_m=210$ $\mu$m.}
\label{fig:restart_c05_xz}
\end{figure}

Figure \ref{fig:restart_c05} shows three snapshots of the normalized concentration field resulting from the original (right half) simulations and those in which the concentration field is restarted right before the downwash (left half). It is evident that, while certain low-concentration patterns in the two boundary layers are nearly identical, others are entirely absent in the latter case. This relatively clear-cut distinction points towards the presence of two separate underlying mechanisms prior and after the time of restart. Furthermore, in figure \ref{fig:restart_c05_xz}, we display a representative snapshot of wall Sherwood number resulting from the two simulations described above. One can clearly recognize that, while the intense-transfer core remains nearly intact in the new simulation, there is no trace of the contact-point peak. It appears that restarting the concentration field right before the identified downwash event does not have a significant influence on the distribution of $\text{Sh}^*$ except for `filtering out' the contact point peak. Apart from that, only minor differences around the rim is observed between the two pictures.

\begin{figure}[H]
\centering

\begin{subfigure}[t]{0.4\textwidth}
    \raggedright $(a)$
    \vspace{0mm}

    \centering
    \raisebox{-0.25cm}{%
        \includegraphics[width=1.25\textwidth]{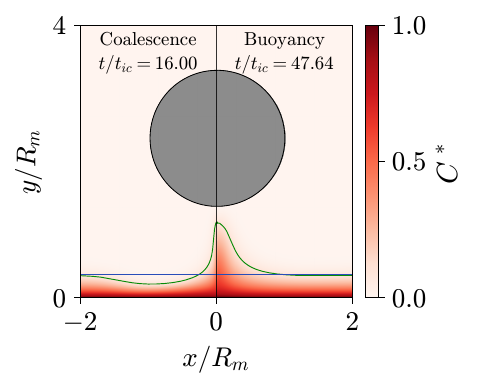}
    }
\end{subfigure}
\hspace{0.08\textwidth}
\begin{subfigure}[t]{0.5\textwidth}
    \raggedright $(b)$
    \vspace{1.5mm}

    \centering
    \includegraphics[width=1.14\textwidth]{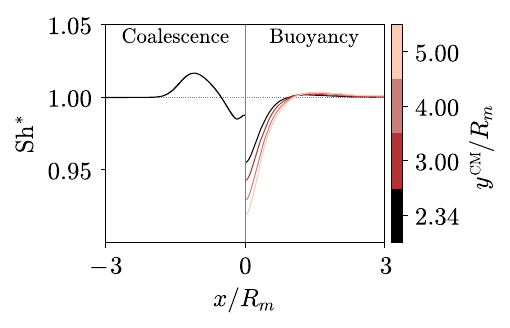}
\end{subfigure}

\caption{Normalized concentration fields (a) and normalized Sherwood numbers (b) for the simulations with concentration field restarted after the bottom of bubbles have risen above the boundary layer edge. The results are shown for the coalescing bubbles (left halves) and purely buoyancy-driven bubble rise (right halves). The contour plots in (a) and the first curve in (b) are at a matched bubble center of mass height $y^{CM}/R_m=2.34$. The green and blue lines in (a) are the iso-contours of $C^*=0.1$ from the simulations and the 1D diffusion solution at the same time, respectively. The simulations are conducted at $\text{Sc}=210$ and $R_m=210$ $\mu$m. }
\label{fig:restart_sh5}
\end{figure}

Before closing this discussion, it is instructive to briefly visit the influence of the second, less strong, micro-convection mechanism, i.e. the one illustrated on the right panel of figure \ref{fig:convection}. Although this mechanism is of secondary importance, a complete picture requires at least an estimate of its contribution. In an attempt to isolate this effect, we run a simulation in which the concentration field is restarted at $t/t_{ic}=2$ -- that is, the time when the south pole of the merged bubble is slightly above $\delta_i$, meaning that the bubble affects the concentration field only indirectly through the induced circulation. The fact that the strong downwash (the first micro-convection mechanism) has already ceased at this point allows separating the effects of these two on the mass transfer coefficient. For reference, the same procedure is applied to the simulation of a bubble rising due to buoyancy starting from a spherical shape at the wall. The results are shown for matched bubble center of mass wall distances in figure \ref{fig:restart_sh5}. Note that this is not a direct comparison between mass transfer of coalescence- and buoyancy-driven bubbles, but a comparison of the strength of a specific micro-convection mechanism given a boundary layer unaffected by any history effects.

It is observed in figure \ref{fig:restart_sh5}(a) that while the general patterns of the two concentration fields are somewhat similar, the exact shape of iso-contours is visibly different. This is evidently a reflection of the fact that, firstly, the coalescence-driven bubble is moderately faster than the buoyancy-driven one, and secondly, the streamlines in the two cases are not completely identical (see figure \ref{fig:axi_3d_vector}). This means that while in both cases the concentration boundary layer thickens below the south pole of the rising bubble, the buoyancy-driven case experiences a much more pronounced thinning around $x/R_m=1$. The impact of these modifications on wall mass transfer can be assessed in figure \ref{fig:restart_sh5}(b), where a small ($\approx2\%$) increase in the Sherwood number is observed at $x/R_m\approx1$. In both cases, the Sherwood number drops in the center due to a thicker boundary layer, although not equally. Note that for the buoyancy-driven case, the less costly axisymmetric simulations have been continued for a longer time. Overall, we can confirm that the effect of the second micro-convection mechanism is minor compared to that of the first. Our numerical experiments at $R_m=150$ $\mu$m, designed to isolate the two, show a maximum local increase in $\text{Sh}^*$ of only a few percent at most due to the second mechanism. This is negligible compared to the nearly two-fold local increase measured due to the first mechanism. The disparity is expected to grow for smaller bubbles since the gap between the two velocity scales ($u_{ic}$ and $u_b$) characterizing the two mechanisms grows when the bubble size drops.

\section{Conclusions}
Interface-resolving numerical simulations are used to study mass transfer on a solid substrate following the coalescence of two equal-sized bubbles. We examine bubbles with radii 25 $\mu$m and 150 $\mu$m, and the physical properties correspond to hydrogen bubbles in water; most importantly, the Schmidt number in liquid phase is $\text{Sc}=210$. Additional simulations at $\text{Sc}=1$ are carried out. The prescribed initial concentration field follows the solution of 1D diffusion into a semi-infinite domain. The initial thickness of concentration boundary layer, defined as the wall distance at which concentration is 10\% of wall concentration, is prescribed to be $1/3R_m$ in the 3D simulations, $R_m$ being the bubble radius. Additional 2D simulations have been carried out to study the effects of intermediate Schmidt numbers and other values of initial boundary-layer thickness.

Immediately following the coalescence, capillary waves rapidly develop on the interface of the merged bubble, eventually leading to a jump-off event. During this period, the bubble center-of-mass velocity and the surrounding liquid velocity scale reasonably well with the inertio-capillary velocity, $u_{ic}$.
This initial phase lasts for less than two inertio-capillary time units ($2t_{ic}$) and is followed by a sharp decrease in bubble rise velocity, which rapidly approaches the terminal velocity of a free-rising bubble of the same size.

The two phases of bubble kinematics are accompanied by two distinct micro-convection patterns, through which low-concentration liquid is entrained into the boundary layer, causing an increase in Sherwood number. Firstly, a strong downward motion is caused directly by the bubble interface velocity in the period of roughly $0.5<t/t_{ic}<1$, i.e., prior to the jump-off event. Secondly, bubble rise after the jump-off causes large-scale circulation in the liquid, through which fresh liquid moves downward to replace the displaced bubble. Notably, the micro-convection due to the first mechanism is much stronger. Note that the Reynolds number based on the characteristic velocity $u_{ic}$ is  $1/\text{Oh}$, which equals 116 and  48 for the larger and smaller bubbles in the present work, respectively. Rapid dissipation of the jump velocity, however, means that the second micro-convection pattern has a smaller characteristic velocity comparable to $u_T$($\sim u_b$).

The computed distributions of concentration and wall Sherwood number indicate that the micro-convection before bubble jump-off can locally elevate mass transfer in a small area contained within the 'shadow' of the merged bubble -- a region we refer to as the enhanced-transfer core. For reference, the computed mean Sherwood number averaged over a circular area of $R_m/2$ radius in time period of $3t_{ic}$ to $16t_{ic}$, equals 1.56 and 1.22 for the larger and smaller bubbles, respectively. It must be kept in mind that the exact amount of increase in $\text{Sh}$ highly depends on the state of the concentration boundary layer around the bubble at the moment of coalescence, which in real-world scenarios can be affected by factors such as current density and proximity of the bubble nucleation sites. The present results indicate that the normalized Sherwood number in the enhanced-transfer core increases with a decrease in the concentration boundary layer thickness. We furthermore study the effect of micro-convection following bubble jump-off. We isolate this effect by artificially ‘restarting’ the concentration boundary layer after the bubble has jumped. This results in only a minor local increase of approximately $2\%$ in the normalized Sherwood number in the bubble wake. Although this increase is larger than that observed for a bubble of the same size rising purely due to buoyancy, it remains negligible compared with the contribution of the first micro-convection mechanism, consistent with the general consensus in the literature.

Finally, it is important to bear in mind that, due to the large computational cost associated with both bubble coalescence and high-Schmidt-number simulations, the present problem is a simplified version of the real-world problem. Most notably, in real-world, the concentration boundary layer at the moment of coalescence is affected by the presence of the bubble. Furthermore, unlike the present simplified boundary condition, wall mass flux is not necessarily uniform, and its determination requires coupling the transport and electric potential equations. Ideally, simulations must include the growth phase of the bubble and extend over detachment time of several bubbles until a dynamic steady state is reached. Realizing such conditions is computationally challenging and can be a subject of future work. What can particularly complicate such simulations is the random timing of coalescence events following random positioning of the nucleation sites in real world. Key insights can, however, be driven from the present simplified set-up. Most importunately, we demonstrate that, unlike the case of pure buoyancy-driven bubble departure, coalescence of two bubbles creates a strong, but short-living, micro-convection event, which depending to the state of the boundary layer can locally modify wall mass transfer. We also show that this effect is very local; the micro-circulation only affects the liquid at the close proximity of the southern hemisphere of the parent bubble, and its potential impact on wall mass transfer does not exceed the area right below the bubble. It is also notable that the flow patterns around coalescing bubbles is significantly different from that of a single bubble on an electrode, whose history effects have been systematically studied in the past \cite{penas2016history,penas2017history}. Different flow patterns modify the concentration boundary layer in different ways, which calls for potential extensions to the understanding of history effects.

\label{sec:conclusion}


\section*{Acknowledgments}
Authors MC and PF acknowledge financial support from Innovation Fund Denmark through the Grand Solution research project LC-H2 (2077-00021B). Authors AS and PF acknowledge financial support from Villum Fonden through the Villum Young Investigator Grant VIL53076. The simulations in this work were conducted using resources provided by the Danish e-Infrastructure Cooperation (project DeiC-AU-N5-2026176).


\section*{Declaration of interests}
The authors report no conflict of interest.

\section*{Data availability}
All data and codes are available upon request.

\vspace{2cm}
\bibliographystyle{jfm}
\bibliography{jfm}


\appendix
\section{Grid--convergence}\label{mesh}

\begin{figure}[H]
\centering
  {\includegraphics[width=1\textwidth]{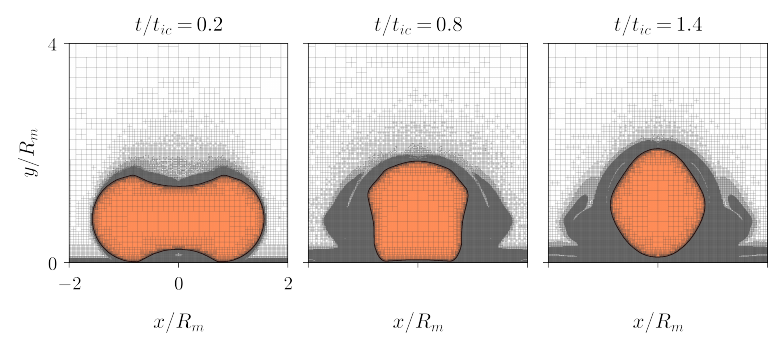}}
  \caption{Adaptive 3D grids for the larger bubble at three representative stages, shown for $\mathrm{Sc}=210$, the most demanding case for scalar transport.}
\label{fig:mesh_3d}
\end{figure}

Figure \ref{fig:mesh_3d} illustrates the adaptive grids used in the three-dimensional simulations at three representative stages of the dynamics: the initial stage of bubble coalescence, before jump-off, and after jump-off. Adaptive refinement is concentrated near the bubble interface, substrate, and wake, where large gradients in concentration fields and flow are expected. For the larger bubble, the minimum grid spacing is $\Delta_{\min}/R_m=0.0058$, corresponding to a maximum refinement level of 12. In contrast, for the smaller one, $\Delta_{\min}/R_m=0.0117$ is used, corresponding to a maximum refinement level of 11. Owing to the adaptive refinement, the number of control volumes varies throughout the simulation, reaching a maximum of approximately $2.36\times10^8$ cells.

Because a systematic refinement study in three dimensions would be computationally prohibitive, the mesh dependence of the solution was first examined systematically in two-dimensional simulations for the larger bubble, which represents the more demanding case in terms of spatial resolution. The two-dimensional calculations employ the same governing equations, adaptive refinement strategy, and wall treatment as the three-dimensional simulations, allowing the spatial-resolution requirements to be assessed systematically at substantially lower computational cost. The resulting convergence analysis was used to establish the required refinement level for the larger-bubble simulations. The selected resolution was subsequently used as a benchmark to assess the required minimum grid size directly in three dimensions.

\begin{figure}[H]
\centering
  {\includegraphics[width=1\textwidth]{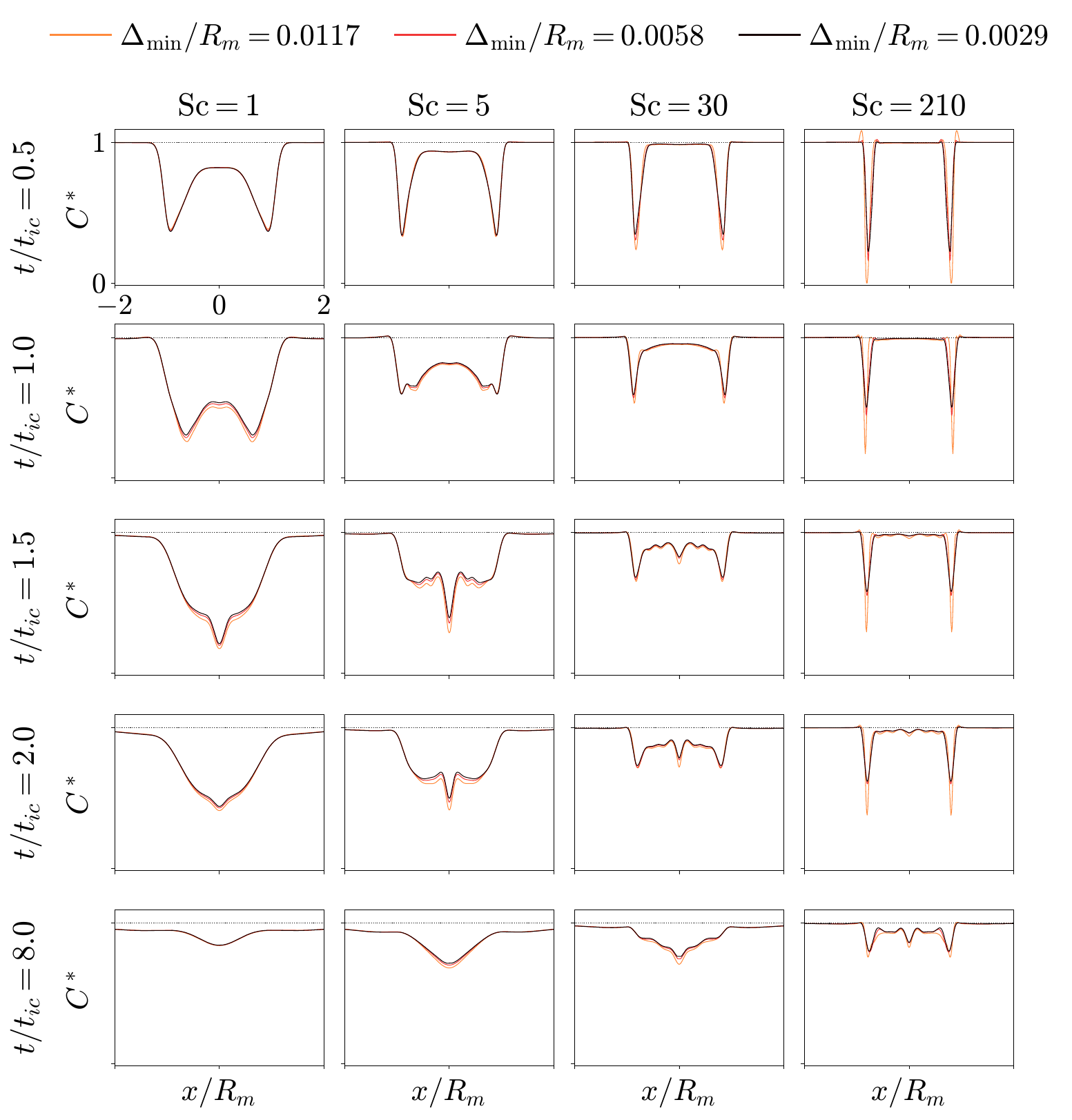}}
  \caption{Normalized wall concentration profiles for the larger bubble at all Schmidt numbers and minimum grid spacings considered in the 2D convergence study. Only the substrate region beneath the bubbles is shown, as the profiles collapse farther away from the bubbles.}
\label{fig:concentration2d_gc}
\end{figure}

\begin{figure}[H]
\centering
  {\includegraphics[width=0.95\textwidth]{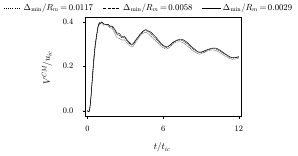}}
  \caption{Evolution of the bubble center-of-mass vertical velocity over time for different minimum grid spacings in 2D.}
\label{fig:vel2d_gc}
\end{figure}

\begin{figure}[H]
\centering
{\includegraphics[width=1\textwidth]{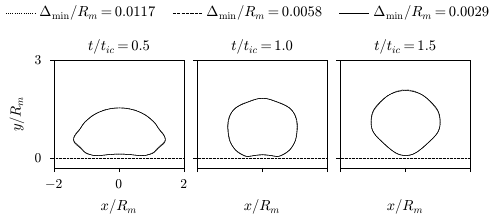}}
  \caption{Bubble interface for different minimum grid spacings.}
\label{fig:shape2d}
\end{figure}

Grid convergence for the 2D simulations was examined for all Schmidt numbers considered in the present study. Three progressively refined meshes with minimum grid spacings $\Delta_{\min}/R_m = 0.0117$, $0.0058$, and $0.0029$ were employed. The convergence analysis was based on the wall concentration profiles beneath the bubbles (figure \ref{fig:concentration2d_gc}), the bubble center-of-mass vertical velocity (figure \ref{fig:vel2d_gc}), and the interface evolution (figure \ref{fig:shape2d}).

The substrate concentration field is the most sensitive quantity to mesh refinement, particularly at large Schmidt numbers, as illustrated in figure \ref{fig:concentration2d_gc}. At high $\mathrm{Sc}$, the scalar diffusivity is small, leading to sharper concentration gradients near the substrate, which require increased spatial resolution. In contrast, for low Schmidt numbers, diffusion smooths the concentration field, resulting in weaker sensitivity to the grid spacing. In the present configuration, the largest discrepancies are observed near the contact-point peaks. Nevertheless, even for $\mathrm{Sc}=210$, the differences between the two finest meshes remain small, indicating satisfactory convergence of the scalar transport solution.

The bubble center-of-mass velocity exhibits weaker sensitivity to the mesh spacing than the concentration profiles (figure \ref{fig:vel2d_gc}), with excellent agreement observed between the two finest grids over the entire time interval considered. This indicates that the hydrodynamic forces governing the bubble motion are sufficiently resolved on the intermediate mesh. The bubble shape displays an even weaker dependence on the grid resolution, with the interface evolution remaining nearly indistinguishable across all meshes throughout the simulations (figure \ref{fig:shape2d}).

Based on the overall agreement obtained for the interface evolution, bubble velocity, and substrate concentration field, the intermediate grid with $\Delta_{\min}/R_m = 0.0058$ was adopted for all 2D simulations reported in the main text, as it provides an excellent compromise between numerical accuracy and computational cost. 

\begin{figure}[H]
\centering
{\includegraphics[width=1\textwidth]{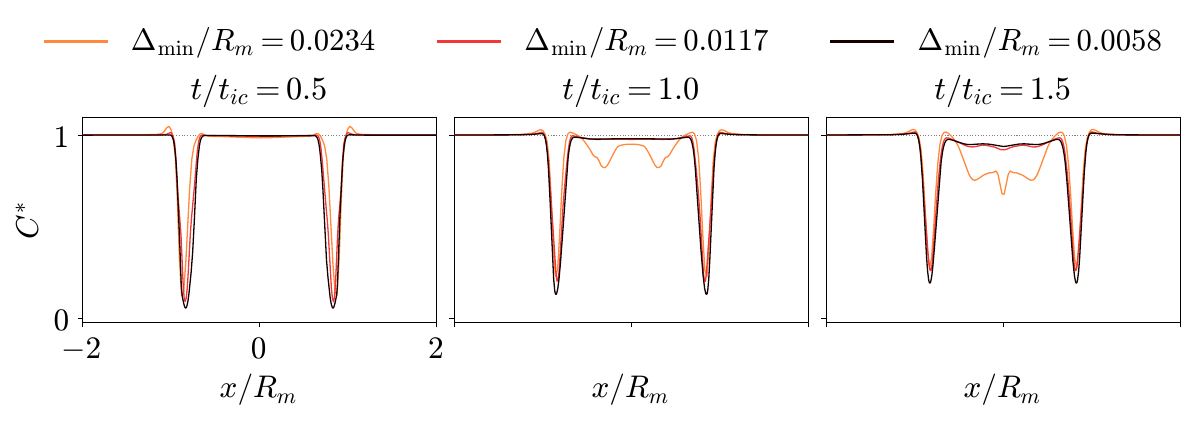}}
  \caption{Normalized wall concentration profiles for the smaller bubble with $\mathrm{Sc}=210$ and minimum grid spacings considered in the 3D convergence study.}
\label{fig:3dsmall}
\end{figure}

The 3D mesh-independence study was conducted for the smaller bubble, as the larger bubble is expected to impose more stringent resolution requirements owing to its more demanding flow field, and hence, is computationally prohibitive. The study was performed at $\mathrm{Sc}=210$, which represents the most demanding case for scalar transport, using three minimum grid spacings, $\Delta_{\min}/R_m=0.0234$, $0.0117$, and $0.0058$. The convergence assessment is based on the normalized wall concentration profiles, shown in figure \ref{fig:3dsmall}. The profiles obtained with $\Delta_{\min}/R_m=0.0117$ and $0.0058$ are essentially indistinguishable, indicating that further refinement has a negligible effect on the scalar transport solution. 
The mesh with $\Delta_{\min}/R_m=0.0117$ is therefore considered sufficient to resolve the smaller-bubble dynamics and was adopted for the corresponding 3D simulations of the smaller bubble.

For the larger bubble, the two-dimensional convergence study identified $\Delta_{\min}/R_m=0.0058$ as an appropriate (and in most snapshots a conservative) choice of mesh mesh, while the three-dimensional mesh-independence study demonstrated that $\Delta_{\min}/R_m=0.0117$ is sufficient for the smaller bubble. Since the larger bubble generates stronger spatial variations in the hydrodynamic field and is therefore expected to impose more stringent spatial-resolution requirements, the finer resolution, $\Delta_{\min}/R_m=0.0058$, was deemed appropriate adopted for its three-dimensional simulations too to ensure adequate resolution.

\section{Domain--size independence}\label{domain}

\begin{figure}[H]
\centering
  {\includegraphics[width=0.725\textwidth]{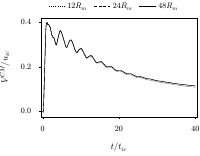}}
  \caption{Evolution of the bubble center-of-mass vertical velocity for different computational domain sizes.}
\label{fig:domain_vby}
\end{figure}

\begin{figure}[H]
\centering
  {\includegraphics[width=1\textwidth]{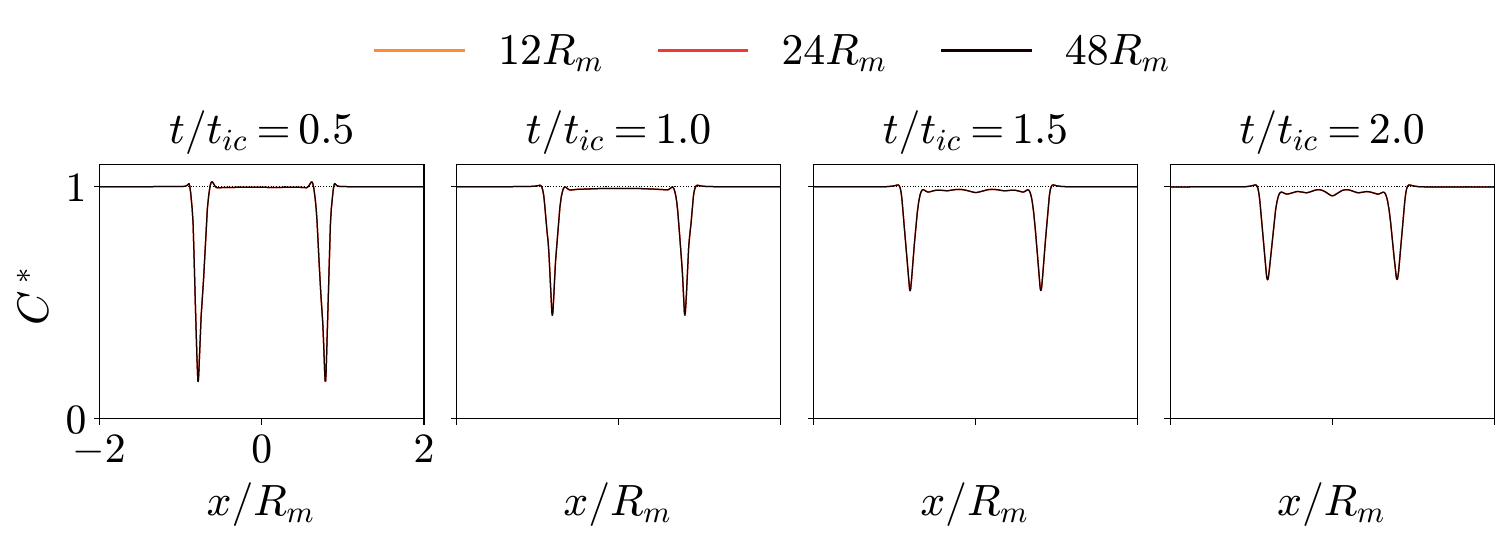}}
  \caption{Normalized wall concentration profiles for different computational domain sizes.}
\label{fig:domain_conc}
\end{figure}

The influence of the computational domain size was assessed in two dimensions using domain sizes $12R_m$, $24R_m$, and $48R_m$. All cases were computed with a mesh resolution of $\Delta_{\min}/R_m = 0.0058$, $\delta_i/R_m=1/3$, and for $\mathrm{Sc}=210$, corresponding to the most demanding scalar-transport conditions considered in the present study. The comparison focuses on the evolution of the bubble center-of-mass vertical velocity (figure \ref{fig:domain_vby}) together with the normalized wall concentration profiles beneath the bubbles (figure \ref{fig:domain_conc}).

Figure \ref{fig:domain_vby} shows that the bubble dynamics are only weakly affected by the computational domain sizes considered. The evolution of the center-of-mass vertical velocity exhibits excellent agreement between the intermediate and largest domains, indicating that confinement effects remain negligible for these configurations. Although the smallest domain also shows good agreement at early times, small deviations become noticeable as the bubbles rise farther from their initial positions and approach the upper boundary of the computational domain. Similarly, the wall concentration profiles shown in figure \ref{fig:domain_conc} display almost no variations between the different domain sizes, demonstrating that the scalar transport remains essentially independent of the computational domain.

Based on the overall agreement obtained for both the bubble dynamics and the wall concentration field, the intermediate domain size, $24R_m$, was adopted for all simulations reported in the present study, as it provides an excellent compromise between minimizing confinement effects and computational cost. The same domain width was also employed as a conservative choice for the three-dimensional simulations as the effect of boundary proximity can be less critical for a three-dimensional domain.

\end{document}